\documentclass[review]{elsarticle}
\usepackage{lineno,hyperref,eurosym}
\usepackage{color}
\usepackage{tabularx}
\usepackage[utf8]{inputenc}
\usepackage{graphicx}
\usepackage{amsmath}
\usepackage{lscape}
\usepackage{xcolor, soul}
\usepackage{makecell}

\journal{Acta Astronautica}
\begin{document}
\begin{frontmatter}
\title{Direct low-energy trajectories to Near-Earth Objects}
\author[label1]{E. Fantino$^{\star}$}
\author[label1]{R. Flores}
\author[label1,label2]{G. Donnarumma}
\author[label3]{D. Canales}
\author[label4]{K.~C. Howell}
\address[label1]{Department of Aerospace Engineering, Khalifa University of Science and Technology, 
P.O. Box 127788, Abu Dhabi (United Arab Emirates)}
\address[label2]{Department of Industrial Engineering, University of Naples Federico II, Piazzale Tecchio 80, 80125, Naples (Italy)}
\address[label3]{Department of Aerospace Engineering, Embry–Riddle Aeronautical University, Daytona Beach, FL 32117 (USA)}
\address[label4]{School of Aeronautics and Astronautics, Purdue University, West Lafayette, IN 47907 (USA)}
\cortext[cor]{Corresponding author: elena.fantino@ku.ac.ae (E.~Fantino)}

\begin{abstract}
Near-Earth Objects (NEOs) are asteroids, comets and meteoroids in heliocentric 
orbits with perihelion below 1.3 au. Similarly to the population of the Main Asteroid Belt, NEOs are primordial bodies and their study can improve our understanding of the origins of the Solar System. With a catalog of over 30~000 known asteroids and approximately 
100 listed short-period comets, the NEO population represents an inventory of exploration targets reachable with 
significantly lower cost than the objects of the Main Asteroid Belt. In addition, the materials present in these bodies could be used to resupply spacecraft en route to other destinations. 
The trajectories of past missions to NEOs have been designed with the patched-conics technique supplemented by 
impulsive and/or low-thrust maneuvers and planetary gravity assist. The transfer times range from some months to a few years, and the close-approach speeds relative to the target have been as high as 10 km/s.  
The design technique described in this work leverages the invariant structures of the circular restricted three-body problem 
(CR3BP) to connect the vicinity of the Earth with NEOs in low-eccentricity, low-inclination orbits. The fundamental building blocks are periodic orbits around the collinear points L$_1$ and L$_2$ of the 
Sun-Earth CR3BP. These orbits are used to generate paths that follow the associated hyperbolic invariant manifolds, exit the sphere of influence of the Earth and reach NEOs on nearby orbits. The strategy is simple, can be applied to depart either a libration point orbit or the vicinity of the Earth, and offers attractive performance features.
\end{abstract}


\end{frontmatter}


\section{Introduction}
\label{sec:introduction}
Near-Earth Objects (NEOs) are asteroids (NEAs) and comets (NECs) with perihelion below 1.3 au \cite{NASA:NEOs}. Aphelia of NEAs generally lie within a sphere of radius 5.2 au, defined by Jupiter's orbit. Only short-period comets (i.e., orbital period less than 200 years) are considered NECs \cite{McFadden:2007}. Scientists believe that the characterization of these objects is key to deepening our understanding of the origin and evolution of the Solar System, and the source of water on Earth. These bodies have also received attention because of their proximity to Earth, which is often associated with collision threats \cite{Anthony:2018, Hedo:2020}. Lastly, in recent years, NEOs have gained importance in the context of resource utilization, as {\it in-situ} collection and storage of material available in space could reduce the cost of space missions \cite{Lewis:1997}. 

The first spacecraft (S/C) to visit a NEO was NASA's NEAR Shoemaker which landed on the surface of 433 Eros in 2001 \cite{Prockter:2002}. This endeavour was followed by NASA's Deep Impact mission to comet Tempel 1 in 2005 \cite{Blume:2005}. The same year, JAXA's Hayabusa probe collected samples from asteroid Itokawa and delivered them to Earth in 2010 \cite{Yoshikawa:2021}. Also in 2010, Deep Impact's mission extension (EPOXI) carried out a flyby with comet Hartley 2 \cite{Chung:2011}. In 2012, CNSA's Change'e 2 executed a close approach with 4179 Toutatis \cite{Liu:2014}. Between 2018 and 2020, JAXA's follow-up mission Hayabusa 2 rendezvoused with asteroid Ryugu, collected samples of its surface and returned them to Earth \cite{Tsuda:2020}. The S/C is now on an extended mission to asteroid 1998 KY26. NASA's OSIRIS-Rex successfully rendezvoused with Bennu, touched down on its surface in 2020 and extracted samples that were delivered to Earth in 2023 \cite{Lauretta:2017, Williams:2018}. The probe is continuing its journey to encounter and study 99942 Apophis in 2029. The Double Asteroid Redirection Test (DART \cite{Rivkin:2023}) by NASA aimed at demonstrating a method for planetary defense. In 2022, DART collided with Dimorphos, a satellite of the asteroid Didymos, and shortened its orbital period by 32 minutes, proving the effectiveness of the transfer of momentum from the S/C to the asteroid. 
The trajectories of these probes were designed using a variety of techniques, including patched-conics, gravity assist, impulsive and continuous-thrust maneuvers. The most relevant performance figures of these missions are summarized in Table~\ref{tab:missions}.

\begin{table}[h!]
\begin{center}
{\scriptsize \begin{tabular}{|r|r|r|r|r|r|r|}\hline \hline
Mission &  Technique & Launch $C_3$ & $\Delta t$ & $\Delta V$ & Outcome & $V_r$ \\ 
        &        &  (km$^2$/s$^2$) & (year) & (m/s) &  & (km/s) \\
\hline \hline
NEAR  \cite{Dunham:2002}           &  GA/PC & 25.9 & 5.0 & 1176  &  SL & 0 \\ \hline
Deep Impact  \cite{Blume:2005}    &  HT & 11.75 & $\sim$0.5 &    -   &   FB+I & 10 \\ \hline
EPOXI   \cite{Chung:2011}         &  GA/PC & extension & 8.0 & $\sim$150    &  FB+I & 12.3  \\ \hline
Hayabusa  \cite{Yoshikawa:2021}       &  LT & - &  2.5   & - & SL & 0.0 \\ \hline
Chang'e 2  \cite{Liu:2014}      &  \makecell{via Moon + \\ L$_2$ Lissajous} &  2.2 & 677 & - & FB & 10.7 \\ \hline
Hayabusa 2  \cite{Tsuda:2020}     &  GA/LT & 21.0  & 3.6 & $\sim$2000 & SL & 0.0 \\ \hline
OSIRIS-Rex \cite{Lauretta:2017}      &  GA/HT & 29.3 & 2.4 & $\sim$900 & SL & 0.0 \\ \hline
DART  \cite{Cheng:2018}           &   HT/LT & - & 0.8 & - & FB+I & 6.0
\\ \hline \hline
\end{tabular}}
\end{center}
{\scriptsize }
\label{tab:missions}
\caption{Trajectory information of past missions to NEOs: transfer technique, launch $C_3$, time of flight to destination ($\Delta t$), $\Delta V$ budget of deterministic maneuvers, target approach mode (outcome) and velocity relative to target at arrival ($V_r$). Meaning of abbreviations and symbols: GA = gravity assist; PC = patched conics; HT = high thrust; LT = low thrust; FB = flyby; SL = soft landing; I = impact; - = data not found.}
\end{table}

The Asteroid Redirect Mission (ARM \cite{ARM:2015}) was a solar-electric propulsion mission proposed by NASA to rendezvous with a large NEA (2011 MD was a prime candidate), use robotic arms to retrieve a 4-meter boulder, and bring it into lunar orbit. In fact, methods to rendezvous, retrieve and deliver portions or entire asteroids to near-Earth space are a major focus of investigations in the context of missions to NEOs. The trajectory design approaches include conventional patched-conics, direct Lambert arcs and gravity assist (GA), and the propulsion can be chemical (high thrust, HT) or electrical (low-thrust, LT) (see, e.g., \cite{Casalino:2002, Santos:2009, Hasnain:2012, Landau:2013, Mascolo:2021}). Notably, Strange et al. \cite{Strange:2010} provide an extensive list of NEAs reachable with HT or LT, directly or via GAs, between 2020 and 2024 and accessible for rendezvous or sample collection and return. Human exploration of NEAs has gained a lot of interest too. Since 2012, NASA has been updating the Near-Earth Object Human Space Flight Accessible Target Study (NHATS) database of targets on orbits very close to Earth's and reachable by a round-trip mission of limited duration \cite{Abell:2012}. The trajectory design is carried out with the method of patched-conics, and employs full-precision ephemerides for both the S/C and the asteroids. Recently, S\'anchez \& Y\'arnoz \cite{Sanchez:2016} proposed a method to drive objects from accessible heliocentric orbits into the Earth's neighborhood using LT and the stable invariant manifolds of Lyapunov and halo orbits. Jorba \& Nicol\'as \cite{Jorba:2021} focused on the capture of NEAs by means of approximations of the stable manifolds of hyperbolic tori associated with the L$_3$ point of the Earth-Moon system in the planar Earth-Moon-Sun bicircular problem.  

\begin{figure}[ht]
\centering
\includegraphics[width=5.0cm]{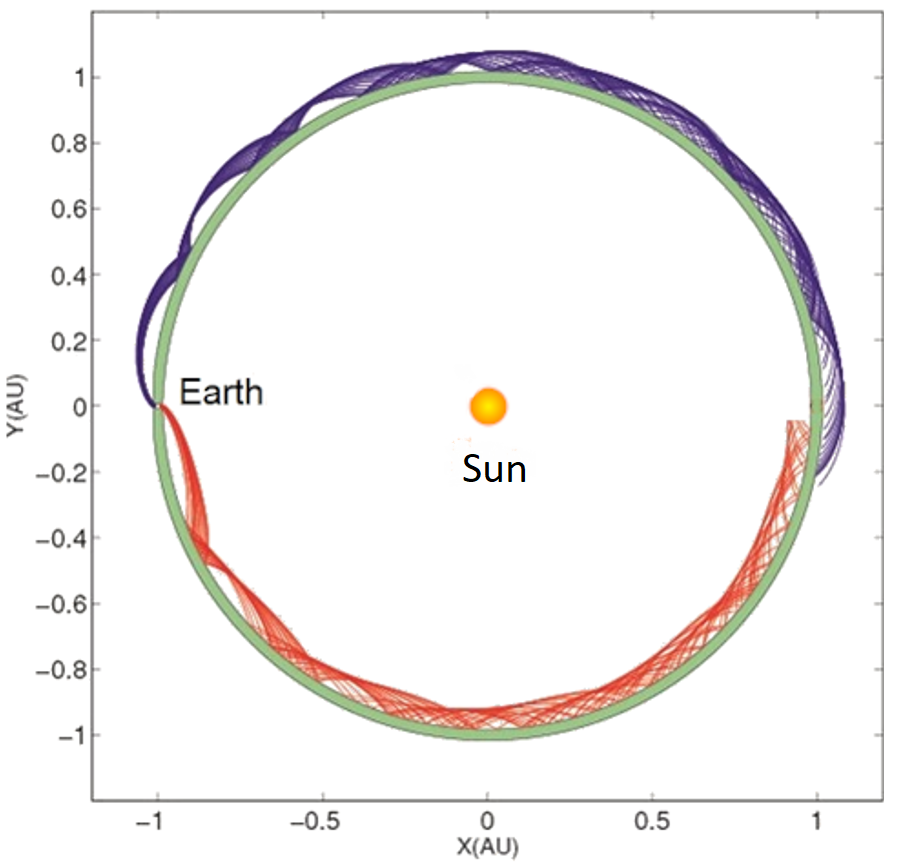} \hspace{1cm} \includegraphics[width=5.0cm]{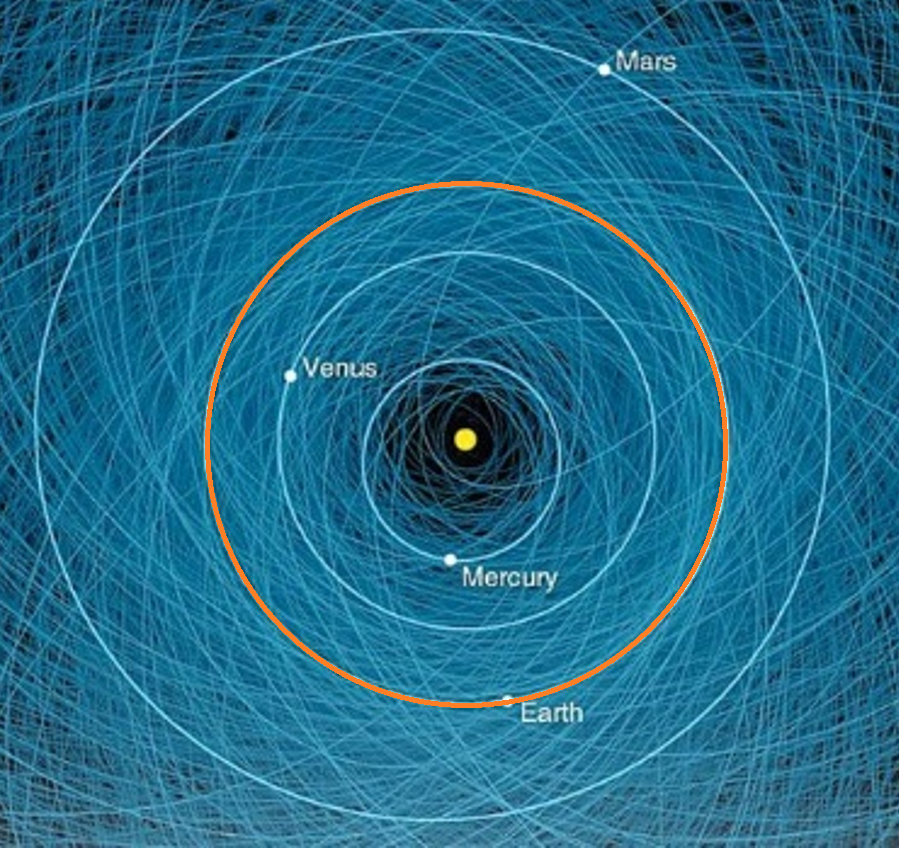} \\
\caption{Left: propagation of branches of hyperbolic invariant manifolds associated with PLOs around L$_1$ (red) and L$_2$ (blue) in the Sun-Earth rotating reference frame (the green ring represents the forbidden region for the specific energy level considered). Right: orbits of several NEOs and the terrestrial planets.}
\label{fig:IMs}
\end{figure}

The use of low-energy trajectories to construct direct transfers and rendezvous opportunities with NEOs has not been addressed in a systematic way. This work offers a simple approach to identify and design this type of solution. The method exploits the dynamical and geometrical features of the hyperbolic invariant manifolds associated with periodic orbits around the collinear equilibrium points L$_1$ and L$_2$ of the Sun-Earth circular restricted three-body problem (CR3BP). These trajectories extend into near-Earth space both inside and outside of the terrestrial orbit. They also acquire a phase difference with respect to Earth's motion, spanning a wide range of orbital domains and coming in close proximity to many resident objects. This is evidenced by Fig.~\ref{fig:IMs}, depicting the paths of unstable hyperbolic invariant manifolds associated with planar Lyapunov orbits (PLOs) around L$_1$ (red) and L$_2$ (blue) in the Sun-Earth rotating reference frame (left), and the orbits of several NEOs together with the terrestrial planets (right).
These natural pathways are characterized by low energies, i.e., low velocities with respect to Earth, which makes them accessible at very low launch cost. A key feature of the method is the approximation of these three-body trajectories with heliocentric two-body (2BP) ellipses beyond a certain distance from Earth. This technique, also called patched CR3BP/2BP model, was developed to design low-energy transfers between Galilean moons, simplifying the computation of connections between invariant manifold trajectories departing and approaching pairs of moons \cite{Fantino:2017, Canales:2021, Canales:2022, Canales:2023}. In a similar way, in this work the encounter between the S/C and a NEO occurs at the intersection between the heliocentric osculating Keplerian orbits of the two bodies, and the rendezvous becomes a simple analytical problem. An impulsive maneuver at an intersection point yields a zero relative velocity encounter between the S/C and the target.
Since the initial conditions of the trajectory of the S/C are generated in the Sun-Earth synodic barycentric reference frame, the orientation in space of the corresponding heliocentric ellipse varies linearly with the departure epoch. This feature provides a degree of freedom enabling the selection of the most desirable solutions, for example on the basis of the $\Delta V$ budget or the transfer time $\Delta t$. 
The methodology was developed in the planar approximation for low-inclination targets, and subsequently extended to 3D solutions.

The paper starts with the definition of the dynamical model for the S/C (Sect.~\ref{sec:model}). Section~\ref{sec:NEOs} deals with the selection of the target NEOs and the processing of their orbital data. Sections~\ref{sec:2D} and \ref{sec:3D} present the methodology and results for planar and 3D transfers, respectively. Section~\ref{sec:comp} compares their performance against conventional patched-conics trajectories. Section~\ref{sec:concl} summarizes and concludes the paper. Preliminary explorations of the method can be found in Donnarumma \cite{Donnarumma:2022}, and have been presented at the 2$^{\rm nd}$ International Stardust Conference \cite{Stardust:2022} and at the $74^{\rm th}$ International Astronautical Congress \cite{CanalesIAC:2023}.

\section{Dynamical model}
\label{sec:model}
\begin{figure}[ht]
\centering
\includegraphics[width=6cm]{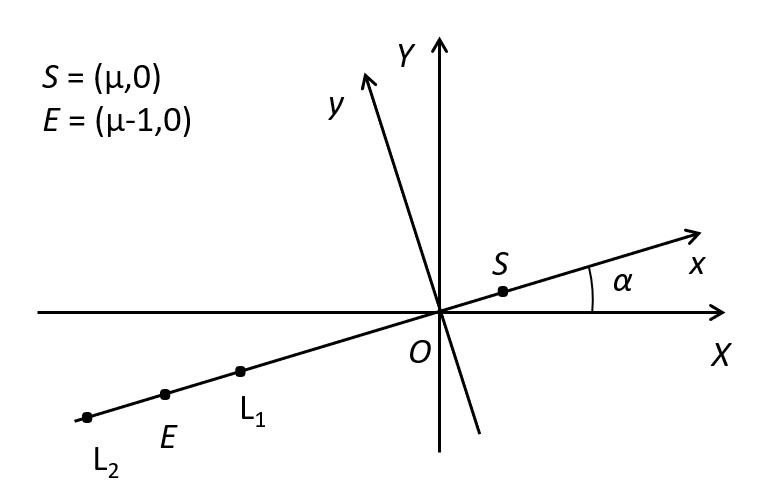} \\
\caption{The Sun-Earth synodic barycentric reference frame ($Oxy$, rotating frame) and the barycentric ecliptic J2000.0 reference frame ($OXY$, inertial frame). $O$ is the barycenter of the Sun-Earth system, $S$ the Sun and $E$ the Earth.}
\label{fig:CR3BP}
\end{figure}

The motion of the S/C starts in the vicinity of the Earth and is governed by the gravitational attraction of the Sun (mass $M_{S}$) and the Earth (mass $M_E$), i.e., the dynamical model is the Sun-Earth-S/C CR3BP. 
If the gravitational constant $G$, the sum $M_{S}+M_E$ and the distance $d$ between the primaries are used as reference magnitudes, the mean motion of the primaries becomes unitary and their orbital period equals $2\pi$. In the synodic barycentric reference frame, the Sun and the Earth are located at at ($\mu$,0,0) and ($\mu-1$,0,0), where $\mu$ denotes the mass ratio $M_E/(M_{S}+M_E)$ of the system (see Fig. ~\ref{fig:CR3BP}). Table~\ref{tab:param} lists the relevant physical parameters of the Sun-Earth system.

\begin{table}[h!]
\begin{center}
\begin{tabular}{|l|l|r|l|}\hline \hline
Symbol & Definition & Value & Units \\ \hline \hline
$GM_{S}$ & Sun's gravitational parameter & $1.3271244 \cdot 10^{11}$ & km$^3$/s$^2$ \\ \hline
$GM_{E}$ & Earth's gravitational parameter & $3.9860044 \cdot 10^{5}$ & km$^3$/s$^2$ \\ \hline
$R_E$ & Earth's equatorial  radius       & 6378.1366                 & km \\ \hline
$\mu$  & Earth-Sun mass ratio      & $0.30034806 \cdot 10^{-5}$ & - \\ \hline
$d$   & Earth-Sun distance         & 149597870              &  km \\ \hline \hline
\end{tabular}
\end{center}
\label{tab:param}
\caption{Physical parameters of the Sun-Earth system used in this work \cite{NASA:constants}.}
\end{table}

\begin{figure}[ht]
\centering
\includegraphics[width=6cm]{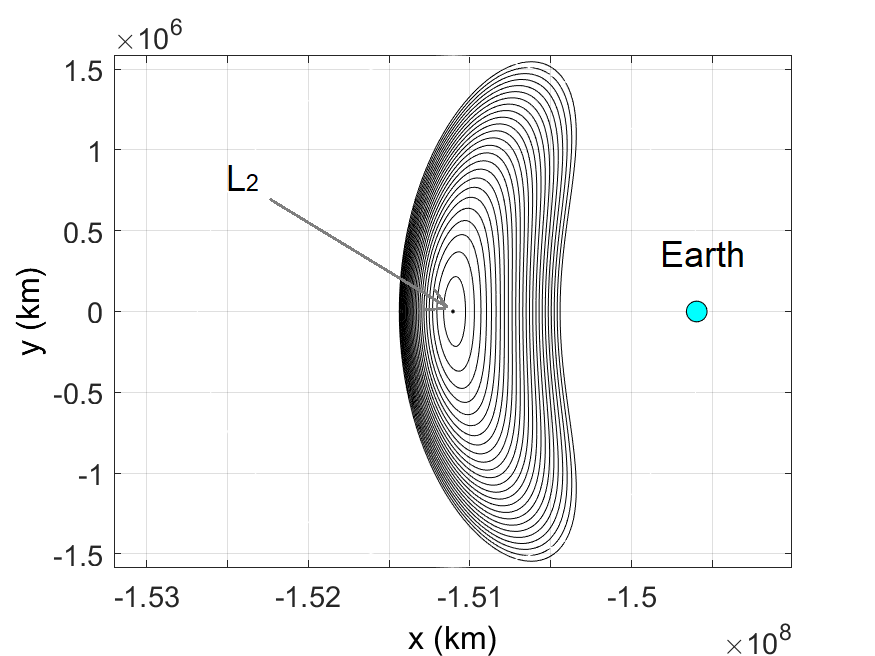} \includegraphics[width=6cm]{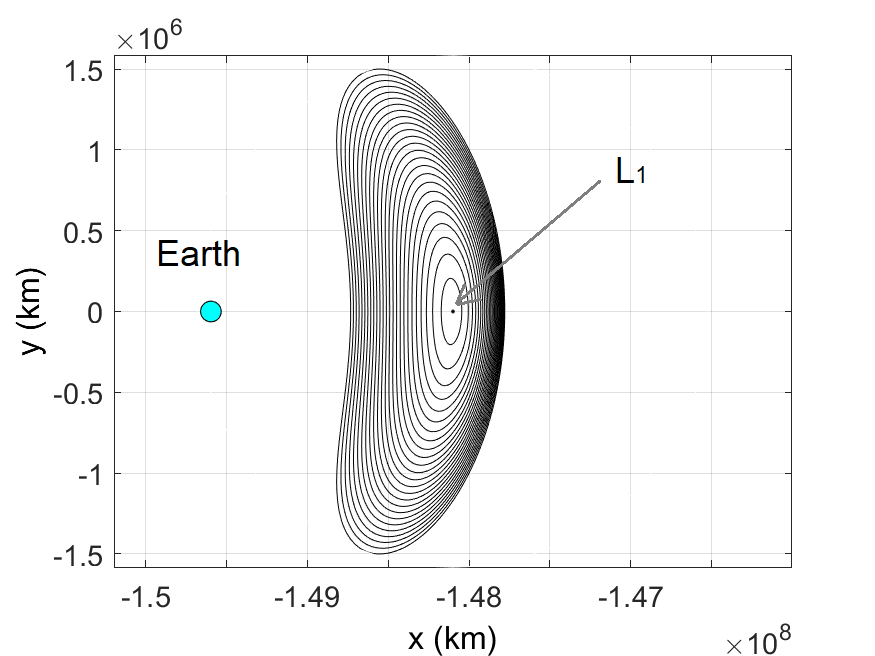} \\
\caption{Families of 50 PLOs around L$_2$ (left) and L$_1$ (right) in the synodic barycentric reference frame.}
\label{fig:PLOs}
\end{figure}

Figure~\ref{fig:PLOs} illustrates families of PLOs around the two collinear equilibrium points L$_1$ and L$_2$. Each set contains 50 orbits equally spaced in Jacobi constant $C_J$ between 3.00056 ($y$ amplitude $\simeq 3 \cdot 10^6$ km) and 3.00089 ($y$ amplitude $\simeq 9 \cdot 10^3$ km).

\begin{figure}[ht]
\centering
\includegraphics[width=6.0cm]{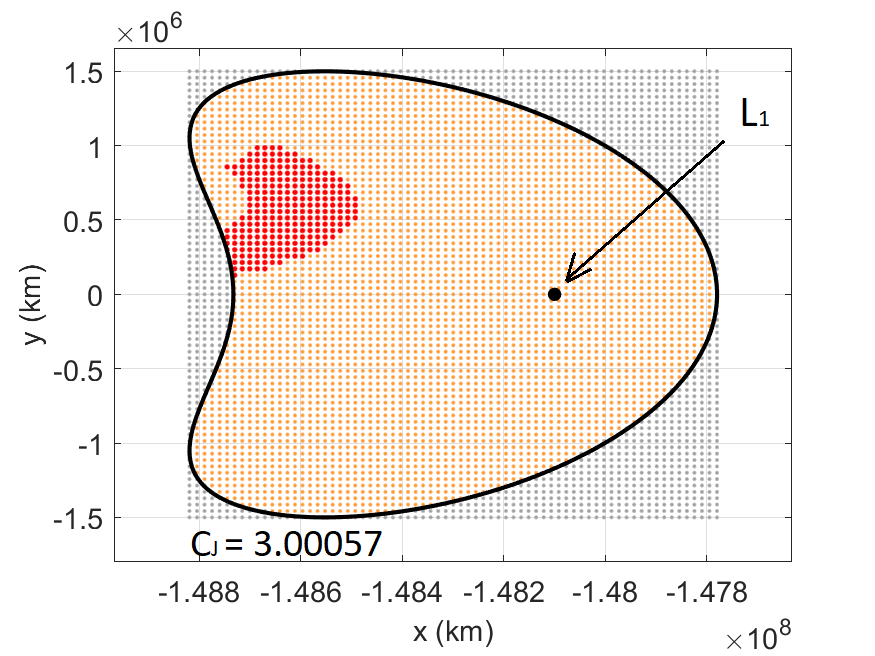} \\
\caption{Generation of initial conditions for TOs inside a PLO with $C_J = 3.00057$: initial 70 x 70 grid (gray), internal points (orange), accepted internal points that connect with a 300-km altitude circular orbit around the Earth (red). For the sake of clarity, the two axes have been drawn with different scales.}
\label{fig:grid}
\end{figure}

\begin{figure}[h!]
\centering
\includegraphics[width=6.0cm]{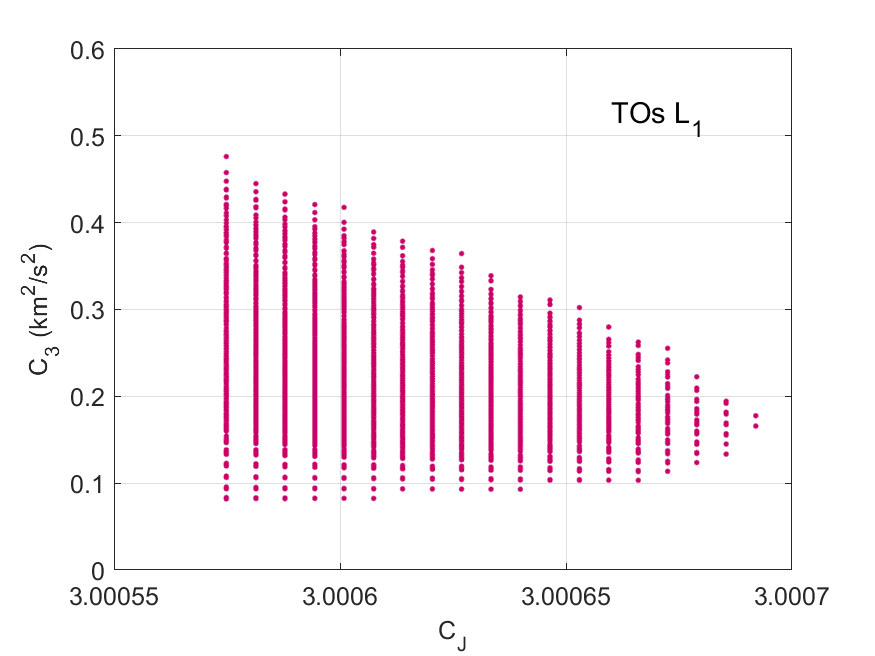}\hspace{-0.15cm} \includegraphics[width=6.0cm]{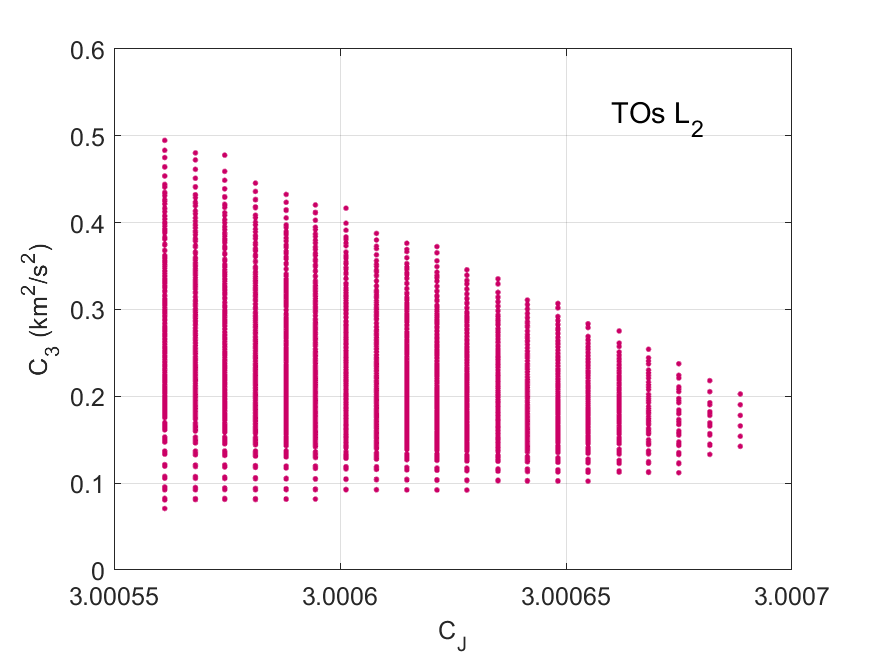} \\
\caption{Characteristic launch energies of the TOs associated with L$_1$ (left) and L$_2$ (right).}
\label{fig:C3}
\end{figure}

\begin{figure}[ht]
\centering
\includegraphics[width=6.0cm]{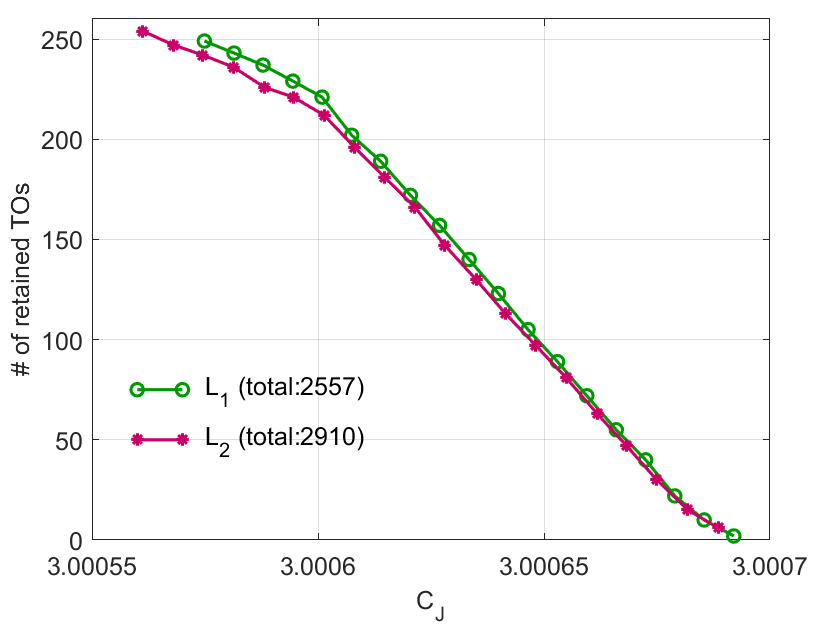} \\
\caption{Number of TOs that reach a 300 km Earth's parking orbit per equilibrium point as a function of $C_J$. 
}
\label{fig:TONum}
\end{figure}

The PLOs serve to generate two types of trajectories:
\begin{itemize}
\item Outward branches of hyperbolic invariant manifolds (MTs) (200 trajectories for each PLO). They are determined and propagated using standard methods. An initial state is generated by applying a small perturbation in the outward direction of the unstable eigenvector of the monodromy matrix of the PLO after appropriate time transformation through the state transition matrix (see e.g., \cite{Parker:1989}), followed by globalization of the manifold by forward time propagation;
\item Transit orbits (TOs), obtained from 70 $\times$ 70 equally-spaced points of a rectangular grid bounding the PLO (Fig.~\ref{fig:grid}). Each point internal to the PLO (orange markers) is the start of a TO with the same $C_J$ as the PLO. The direction of the initial velocity is parallel (L$_1$) or anti-parallel (L$_2$) to the $x$-axis\footnote{Other choices are possible because the position and the Jacobi constant only constrain the magnitude of the velocity.}. In this way, the TOs drive the S/C away from the Earth forward in time along paths contained in the outward invariant manifold branch with the same $C_J$. The initial states of the TOs are propagated backwards in time (i.e., towards the Earth). The trajectories that cross a circular parking orbit at 300 km altitude are retained (red markers in Fig.~\ref{fig:grid}), the others are discarded. 
Figure~\ref{fig:C3} shows the characteristic launch energies ($C_3$) of the retained TOs, computed in the Earth-S/C 2BP. 
The number of retained TOs varies with the amplitude (hence, the value of $C_J$) of the PLO. Only the first 20 Lyapunov orbits of each family provide solutions that reach Earth's vicinity (Fig.~\ref{fig:TONum}). 
\end{itemize}

\begin{figure}[ht]
\centering
\includegraphics[width=5.5cm]{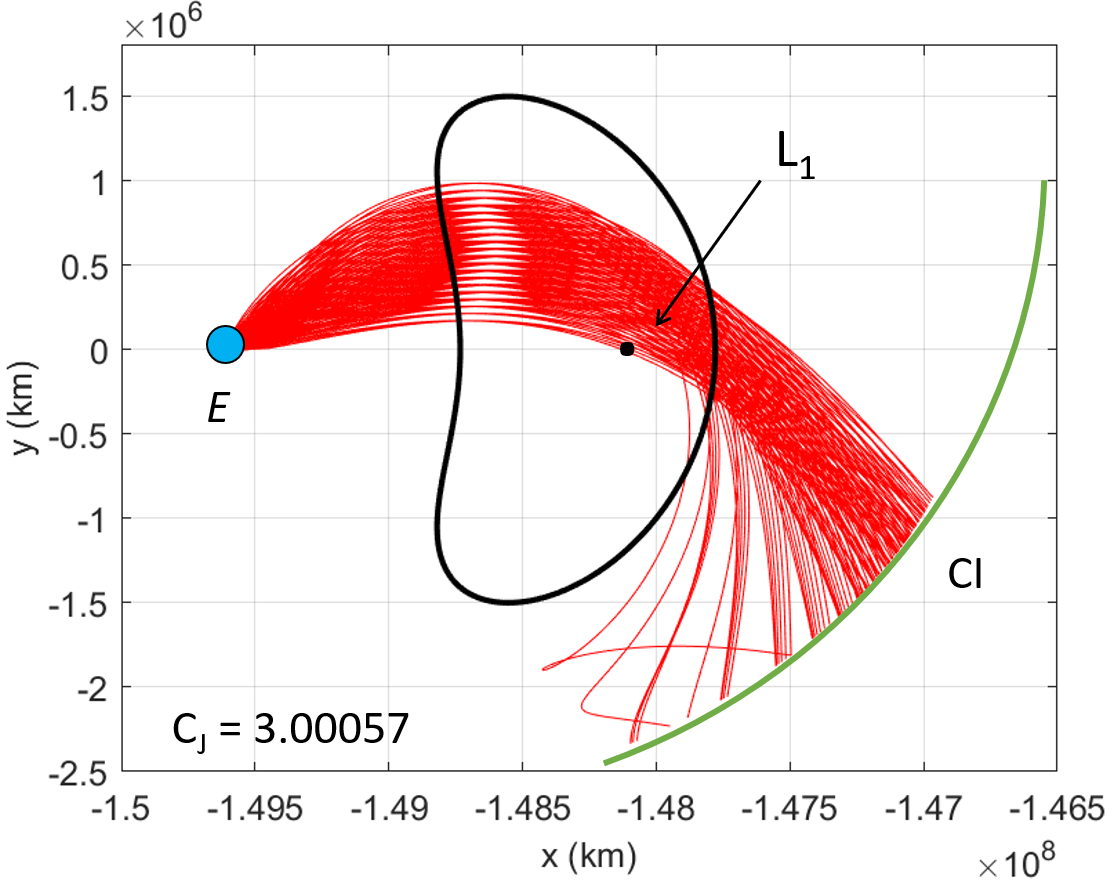}  \includegraphics[width=6cm]{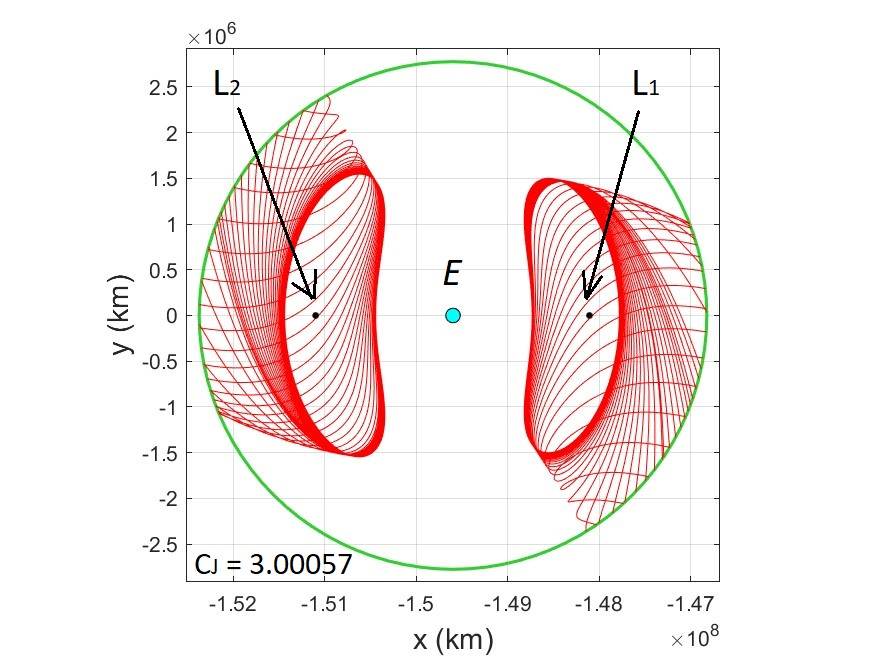}\\
\caption{Left: propagation of TOs from the the accepted grid points towards the parking orbit (backward) and the CI (forward). Right: forward propagation of the outward branches of two MTs until they cross the CI.}
\label{fig:CI}
\end{figure}

\begin{figure}[ht]
\centering
\includegraphics[width=6.5cm]{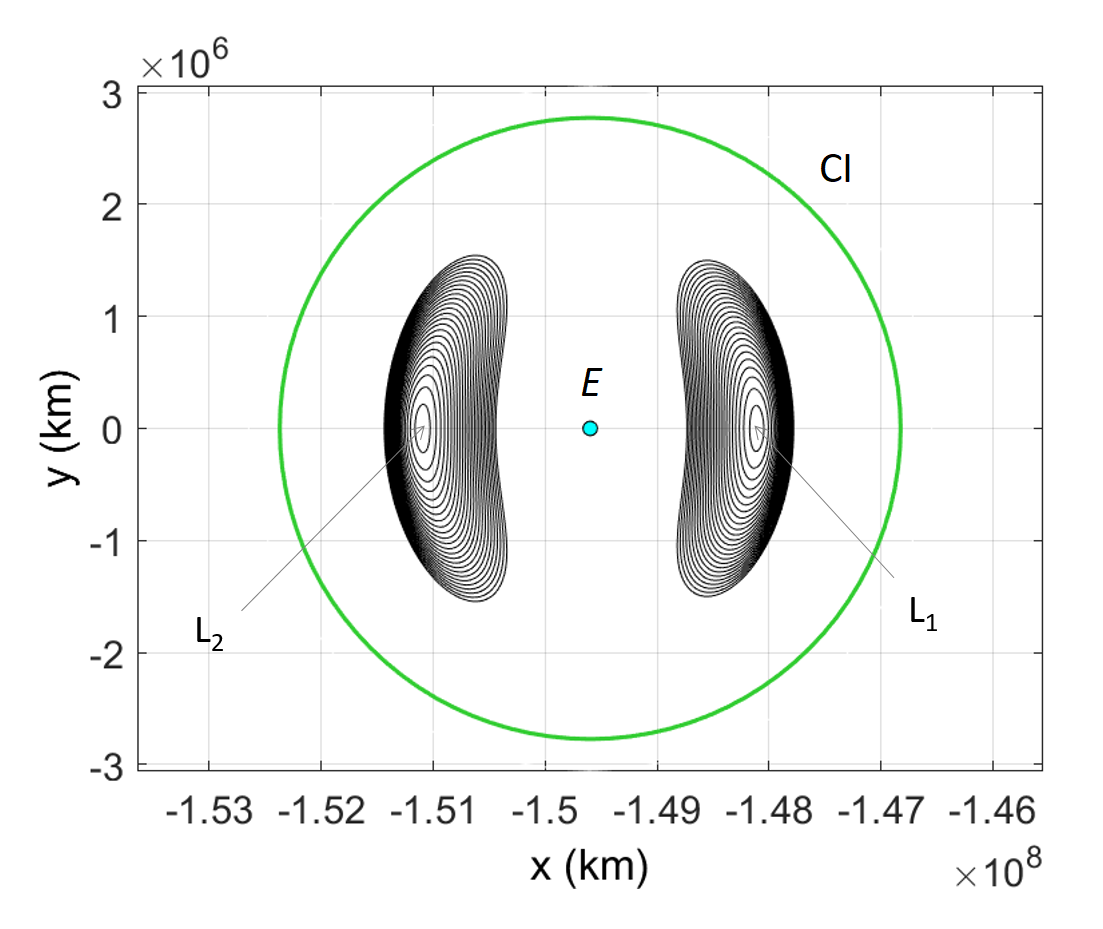} \\
\caption{The two families of PLOs and the CI.}
\label{fig:PLO_CI}
\end{figure}

The accepted TOs and all the MTs are propagated forward in time until they intersect an Earth-centered circle, called circle of influence (CI, Fig.~\ref{fig:CI}) whose radius $r_{CI}$ equals 2~773~940 km or three times $r_{SoI} = d \; (M_E/M_{S})^{2/5}$, the radius of the Laplace sphere of influence of the Earth. This choice ensures that the CI encloses all the PLOs of the two families, as shown in Fig.~\ref{fig:PLO_CI}.  

\begin{figure}[h!]
\centering
\includegraphics[width=6.cm]{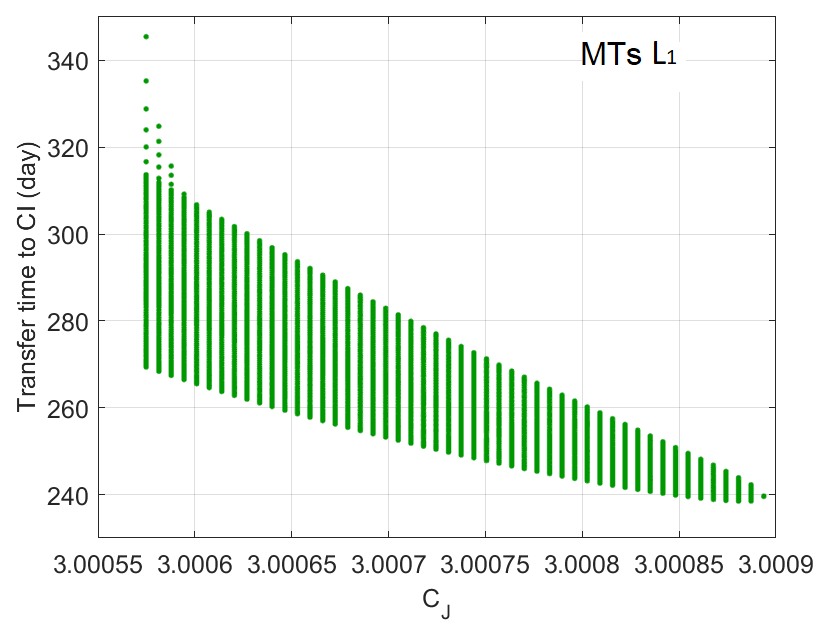} \includegraphics[width=5.95cm]{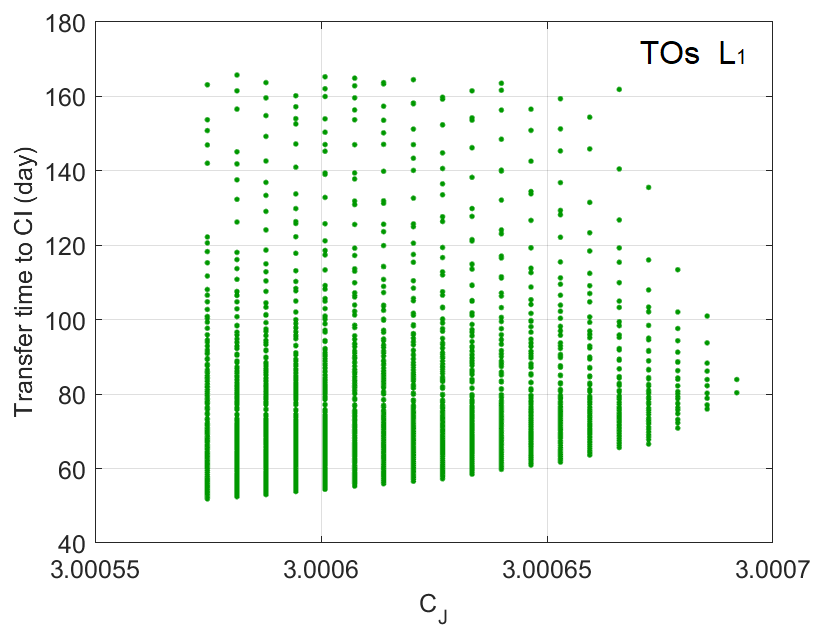}\\
\includegraphics[width=6.cm]{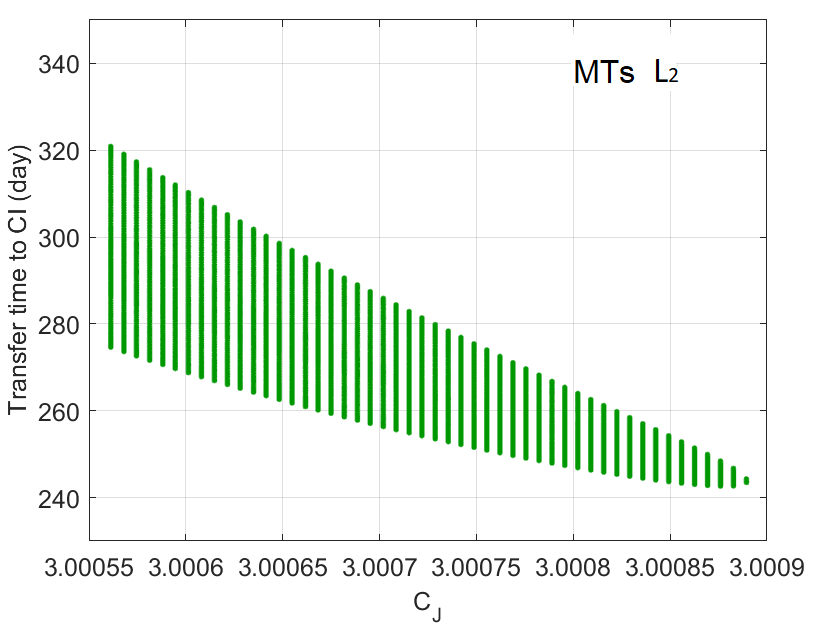} \includegraphics[width=6.cm]{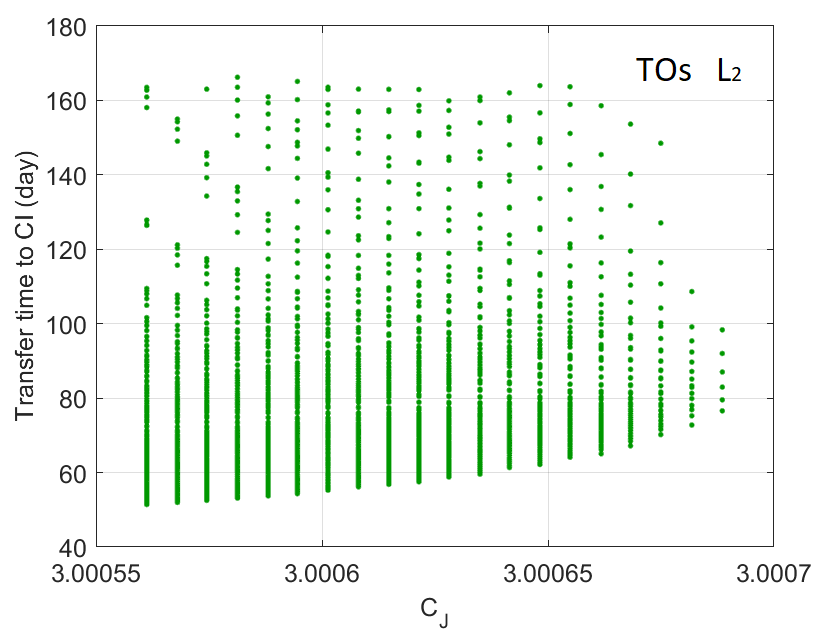}\\
\caption{Transfer times from each departure location to the CI for each trajectory type.}
\label{fig:TOF}
\end{figure}

MTs and TOs represent different mission opportunities. MTs can be used, for example, to depart the vicinity of an equilibrium point, whereas TOs are direct trajectories from a geocentric parking orbit. The two types span the same energy levels, but while MTs initially stay close to the LPO, TOs pass through it and reach the CI directly, resulting in faster transfers (see Fig.~\ref{fig:TOF}).


\begin{figure}[ht]
\centering
\includegraphics[width=8cm]{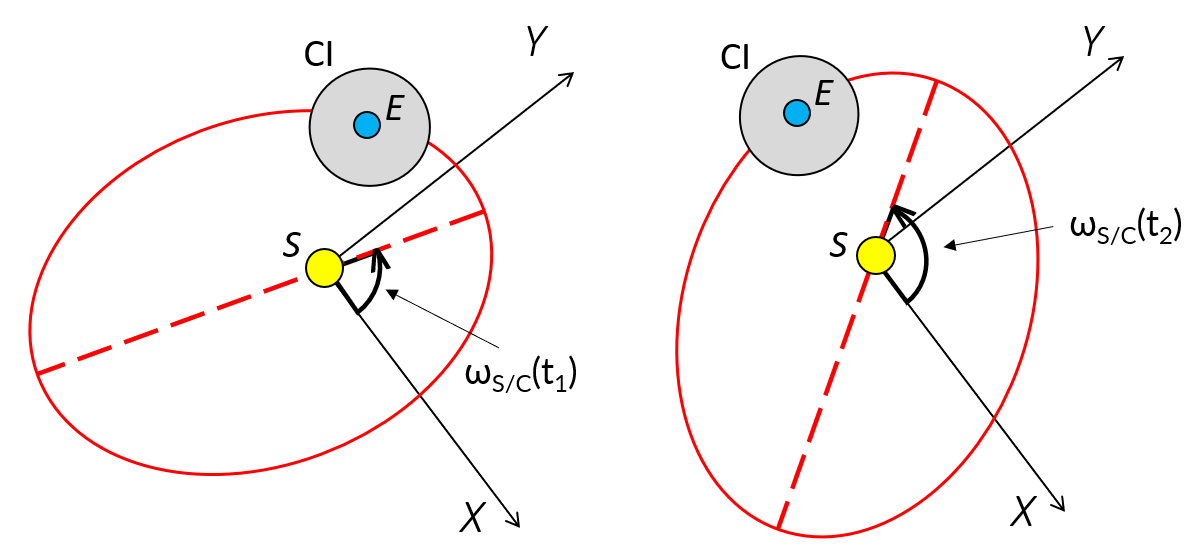} \\
\caption{Effect of the CI crossing time ($t_1 < t_2$) on the longitude of the perihelion $\omega_{S/C}$ of an osculating heliocentric orbit obtained from the same CR3BP state vector at the CI.}
\label{fig:omega_rot}
\end{figure}

The state vectors of MTs and TOs at the CI are collected and transformed to the heliocentric ecliptic J2000.0 reference frame. In this operation, the distance between the barycenter of the Sun-Earth system and the center of the Sun is neglected. The longitude of the perihelion of the osculating heliocentric ellipses depends on the date of CI crossing, and varies at a constant rate equal to the mean motion $n_{ES}$ of the Sun-Earth system. Semimajor axis, eccentricity and true anomaly depend entirely on the state vector at the CI in the synodic frame. This is sketched in Fig.~\ref{fig:omega_rot} which shows two osculating ellipses corresponding to the same three-body state vector at two different CI crossing times, $t_1$ and $t_2$ (resulting from two different departure dates on the same CR3BP trajectory). The shapes of the two orbits are identical, and they share the same true anomaly at CI crossing. The directions of the lines of apsides are related through $\omega_{S/C}(t_2) = \omega_{S/C}(t_1) + n_{ES} (t_2 - t_1)$, where $\omega_{S/C}$ is the longitude of the perihelion of the S/C's orbit. The value of this parameter depends on the angle ($\alpha$ in Fig.~\ref{fig:CR3BP}) between the synodic frame and the inertial frame at the time of CI crossing. This angle is zero at the Spring Equinox, whose reference epoch is 2460024.3917 JD = 2023-Mar-20 21:24:00 UTC. This date is used to determine the instantaneous orientation of the $x$-axis of the synodic frame. 

\section{NEOs: orbital data and dynamical model}
\label{sec:NEOs}
\begin{figure}[h!]
\centering
\includegraphics[width=5.5cm]{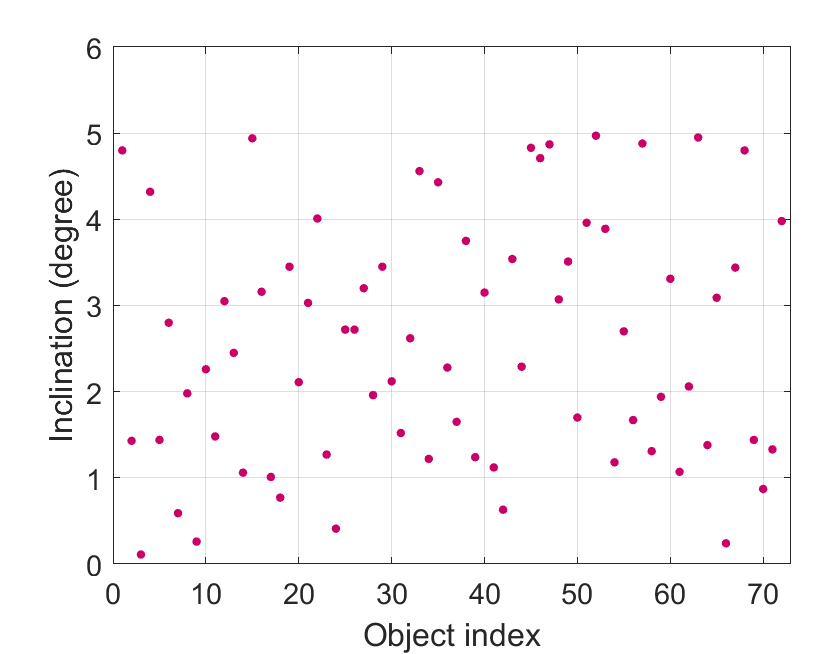} 
\includegraphics[width=5.5cm]{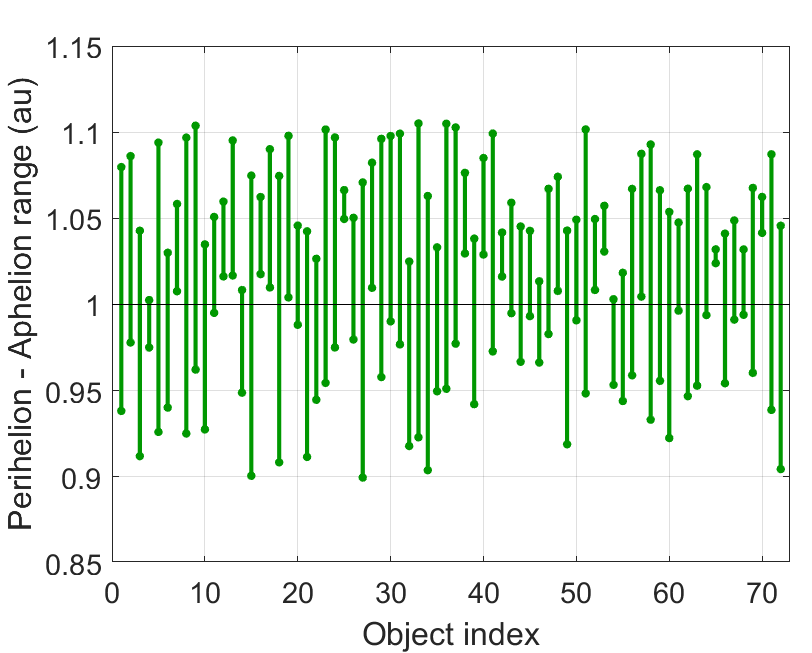} \\
\caption{Orbital inclinations (left) and perihelion - aphelion ranges (right) of the 72 selected NEOs.}
\label{fig:NEOs}
\end{figure}

The orbital parameters of the candidate targets for rendezvous have been extracted from the Solar System Dynamics Minor Bodies Database \cite{NASA:small_bodies}, which on July 30, 2023 listed orbital data for 32~396 asteroids and comets with perihelion below 1.3 au. Retaining only objects with ecliptic inclination $\le$ 5$^{\circ}$, perihelion distance $\ge$ 0.9 au and aphelion distance $\le$ 1.1 au yields a set of 72 asteroids (Table~\ref{tab:design}). The semimajor axes range from 0.971 to 1.058 au, while the eccentricities span the interval [0.039, 0.0898]. Figure~\ref{fig:NEOs} shows the distribution of orbital inclinations and the intervals of heliocentric distances swept by the 72 orbits. All data are given in the heliocentric ecliptic J2000.0 reference frame with osculation epoch $t_0$ = 2460000.5 JD = 2023-Feb-25 00:00:00 UTC.
The dynamical model for the NEOs is the Sun-asteroid 2BP.
\begin{table}[h!]
\begin{center}
{\scriptsize \begin{tabular}{|l|l|l|l|l|}
 \hline \hline
   ~1 =   459872        & 16  = 2012 LA           & 31  = 2017 FT102	  &  46  = 2020 HO5  & 61 =  2022 NX1  \\ \hline   
   ~2 =  1991 VG        & 17  = 2012 TF79 	 & 32  = 2017 HU49 	  &  47  = 2020 MU1  & 62 =  2022 OB5  \\  \hline
   ~3 =  2000 SG344     & 18  = 2013 BS45 	 & 33  = 2018 FM3  	  &  48  = 2020 RB4  & 63 =  2022 RS1  \\  \hline
   ~4 =  2003 YN107     & 19  = 2013 GH66 	 & 34  = 2018 PK21 	  &  49  = 2020 VN1  & 64 =  2022 RD2  \\  \hline
   ~5 =  2006 JY26      & 20  = 2013 RZ53 	 & 35  = 2018 PN22 	  &  50  = 2020 WY   & 65 =  2022 RW3  \\  \hline
   ~6 =  2006 QQ56      & 21  = 2014 DJ80 	 & 36  = 2018 PM28 	  &  51  = 2021 AT2  & 66 =  2023 FY3  \\  \hline
   ~7 =  2006 RH120     & 22  = 2014 QD364	 & 37  = 2018 WV1 	  &  52  = 2021 AK5  & 67 =  2023 GQ1  \\  \hline
   ~8 =  2008 KT        & 23  = 2014 WU200	 & 38  = 2019 FV2 	  &  53  = 2021 CZ4  & 68 =  2023 GT1 \\  \hline
   ~9 =  2008 UA202     & 24  = 2014 WX202	 & 39  = 2019 GF1 	  &  54  = 2021 GM1  & 69 =  2023 HM4   \\ \hline
  10 =  2010 JW34      & 25  = 2015 JD3  	 & 40  = 2019 KJ2 	  &  55  = 2021 LF6  & 70 =  2023 HG11  \\ \hline
  11 =  2010 VQ98      & 26  = 2015 XZ378 	 & 41  = 2019 PO1 	  &  56  = 2021 RZ3  & 71 =  2023 LE    \\ \hline
  12 =  2011 BL45      & 27  = 2016 GK135 	 & 42  = 2020 CD3 	  &  57  = 2021 RG12 & 72 =  2023 LG2    \\ \hline
  13 = 2011 MD         & 28  = 2016 RD34  	 & 43  = 2020 FA1         &  58  = 2021 VH2  & 	    	      \\ \hline
  14 = 2011 UD21       & 29  = 2016 YR    	 & 44  = 2020 GE          &  59  = 2021 VX22 &		      \\ \hline
  15  = 2012 FC71      & 30  = 2017 BN93 	 &  45  = 2020 HF4	  &  60  = 2022 BY39 &		      \\ 
  \hline \hline
\end{tabular}}
\end{center}
\caption{Candidate targets: primary designation and object index assigned in this work.}
\label{tab:design} 
\end{table}

\section{Planar impulsive rendezvous}
\label{sec:2D}
\begin{figure}[h!]
\centering
\includegraphics[width=11cm]{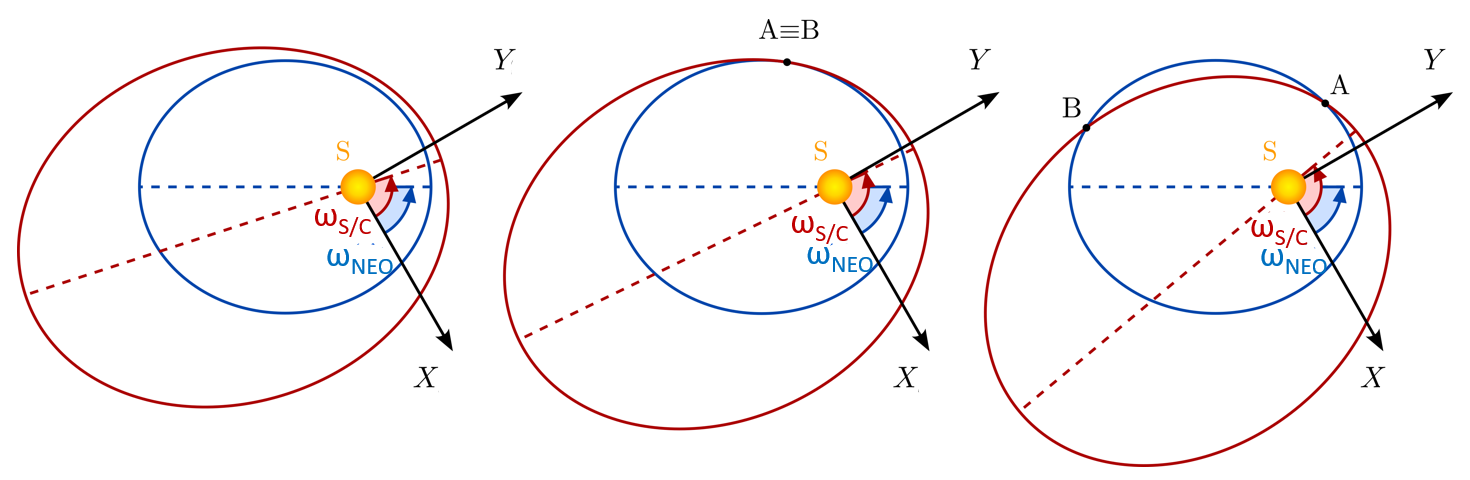} \\
\caption{Number of intersection points between confocal coplanar ellipses for different mutual orientations: 0 (left), 1 (center), 2 (right).}
\label{fig:Intersect}
\end{figure}

Neglecting the gravitational attraction of the Earth outside the CI transforms the design of a direct transfer to a NEO into a search for intersections between confocal ellipses. In the planar approximation, the selected low-inclination targets and the S/C move in the ecliptic plane. Hence, the rendezvous occurs at the intersection between their coplanar confocal elliptical orbits. The solution of this mathematical problem is due to \cite{Wen:1961} and was recently applied to design transfers between giant planet moons \cite{Fantino:2017, Canales:2021, Canales:2022, Canales:2023}. The semimajor axis and the eccentricity of the orbits of the two bodies, and the angle between the two periapses determine the number of intersection points. As shown in Fig.~\ref{fig:Intersect}, which portrays two elliptical orbits with their arguments of perihelion $\omega_{NEO}$ and $\omega_{S/C}$, the number of intersection points can be 0, 1 (tangential intersections A = B) or 2 (A $\neq$ B) depending on the relative sizes and orientations of the ellipses. The tangential case corresponds to the rendezvous impulse $\Delta {\bf V}_{\#} = {\bf V}_{S/C} - {\bf V}_{NEO}$ with the smallest magnitude (see, e.g., \cite{Fantino:2017}).

\begin{figure}[h!]
\centering
\includegraphics[width=12cm]{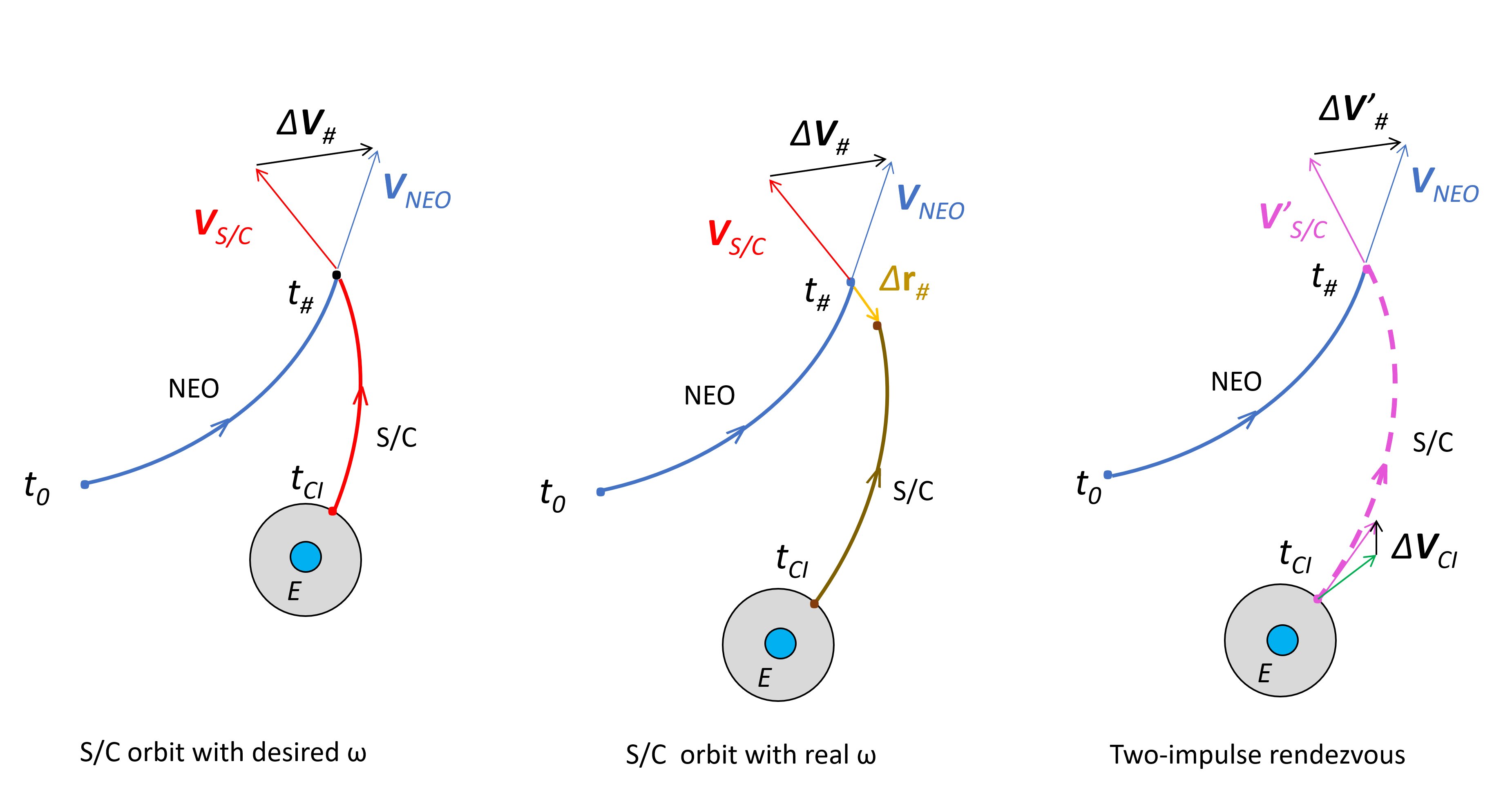}\\
\caption{Left: Ideal S/C transfer and rendezvous with $\omega _{S/C} = \omega_d$. Middle: real S/C transfer with $\omega _{S/C} = \omega_r$. Right: Two-impulse transfer through the solution of Lambert's problem between $t_{CI}$ and $t_{\#}$.}
\label{fig:schemes}
\end{figure}

The longitude of the orbit perihelion for the NEO is fixed at the value corresponding to the osculation epoch. For the S/C this quantity is a function of the CI crossing date, which ultimately depends on the departure date. The latter is computed {\it a posteriori} on the basis of the rendezvous requirements, as shown later\footnote{The autonomous character of the CR3BP ensures freedom in the choice of the departure date.}.
The range of values of the longitude of the perihelion of the orbit of the S/C for which intersections with the target orbit exist can be determined analytically (see \cite{Fantino:2017}). This range is sampled with a uniform step $\Delta \omega$, and each value (denoted $\omega_d$ meaning desired $\omega$) is used to compute the position of the intersection point(s) between the elliptical orbits. The time of passage of the target NEO through the rendezvous point determines the rendezvous date $t_{\#}$ (Fig.~\ref{fig:schemes} left). From this epoch, the (desired) time $t_{CI}$ of CI crossing is computed. This date is associated with a value (denoted $\omega_r$ or real $\omega$) of the longitude of the perihelion of the S/C orbit. A non-zero $\omega_{e}$ = $\mid \omega_d - \omega_r \mid$ yields a non-zero rendezvous distance $\Delta r_{\#}$ between the S/C and the target at $t_{\#}$ (Fig.~\ref{fig:schemes} middle). $\Delta r_{\#}$ depends on $\omega_e$ and the eccentricity of the two orbits. In the limit case of zero eccentricity and orbital radius = $r$ for both the S/C and the target, $\Delta r_{\#} = 2r \sin(\omega_e/2)$, which for $\omega_e$ = 0.01$^{\circ}$ and $r$ = 1 au gives $\Delta r_{\#} \simeq 10^4$ km. Therefore, limiting the value of $\omega_e$ (by discarding solutions for which this quantity exceeds a pre-assigned threshold) leads to  encounters with an acceptable close-approach distance $\Delta r_{\#}$. Then, $\Delta {V}_{\#}$ and the total transfer time $\Delta t$ can be used to characterize the performance of a rendezvous trajectory with the given target.  

Note that $\Delta r_{\#}$ can be reduced to zero, for instance, by computing the  Lambert arc between the position of the S/C at $t_{CI}$ and that of the asteroid at $t_{\#}$, leading to a two-impulse transfer, i.e., $\Delta {\bf V}_{CI}$ at the CI and $\Delta {\bf V}_{\#}'$ at rendezvous (Fig.~\ref{fig:schemes} right). 
Alternatively, $\Delta r_{\#}$ can be reduced or eliminated by replacing the impulsive rendezvous maneuver ($\Delta {\bf V}_{\#}$) with a LT transfer.

\begin{figure}[h!]
\centering
\includegraphics[width=6.1cm]{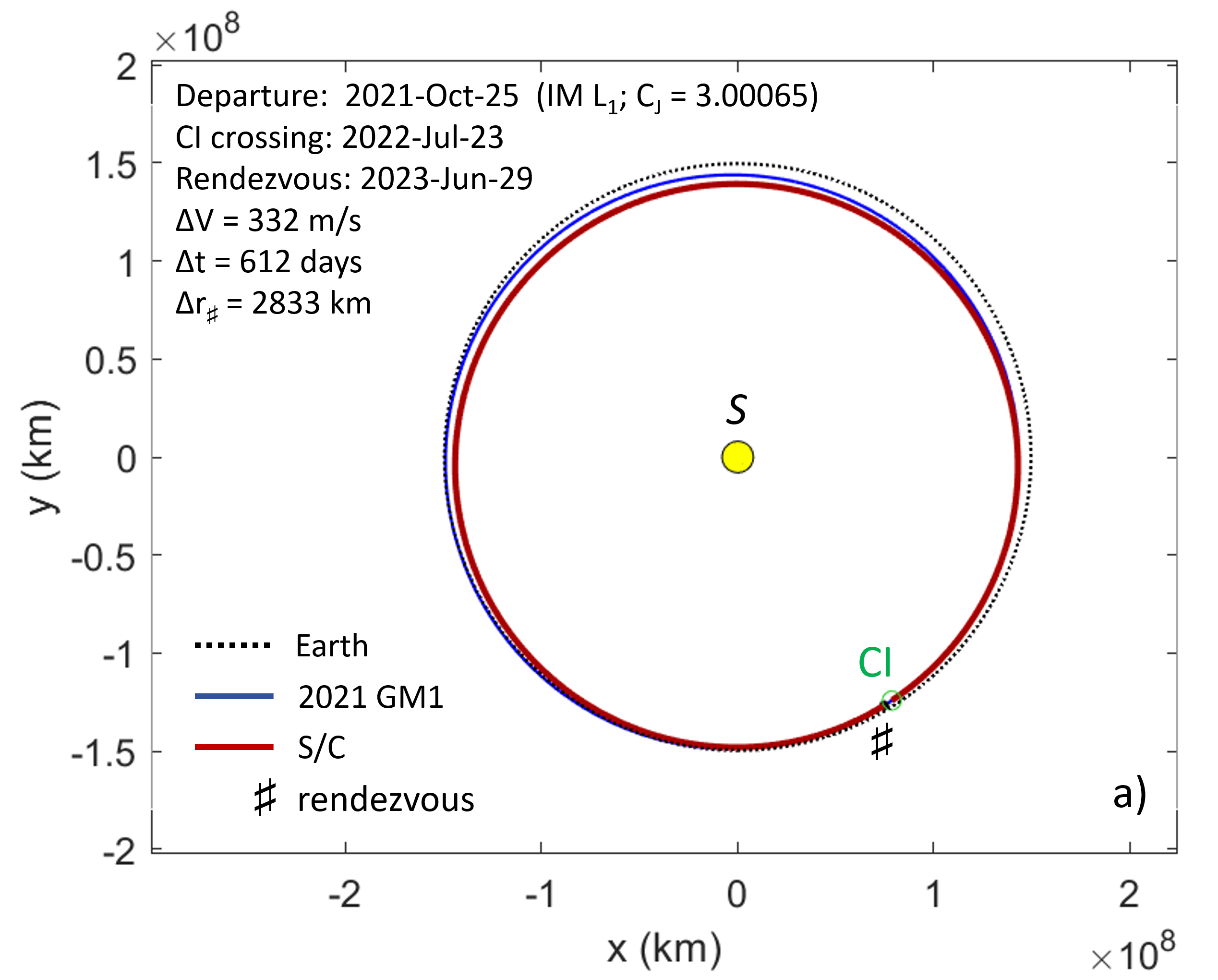} \hspace{-0.3cm} \includegraphics[width=6.1cm]{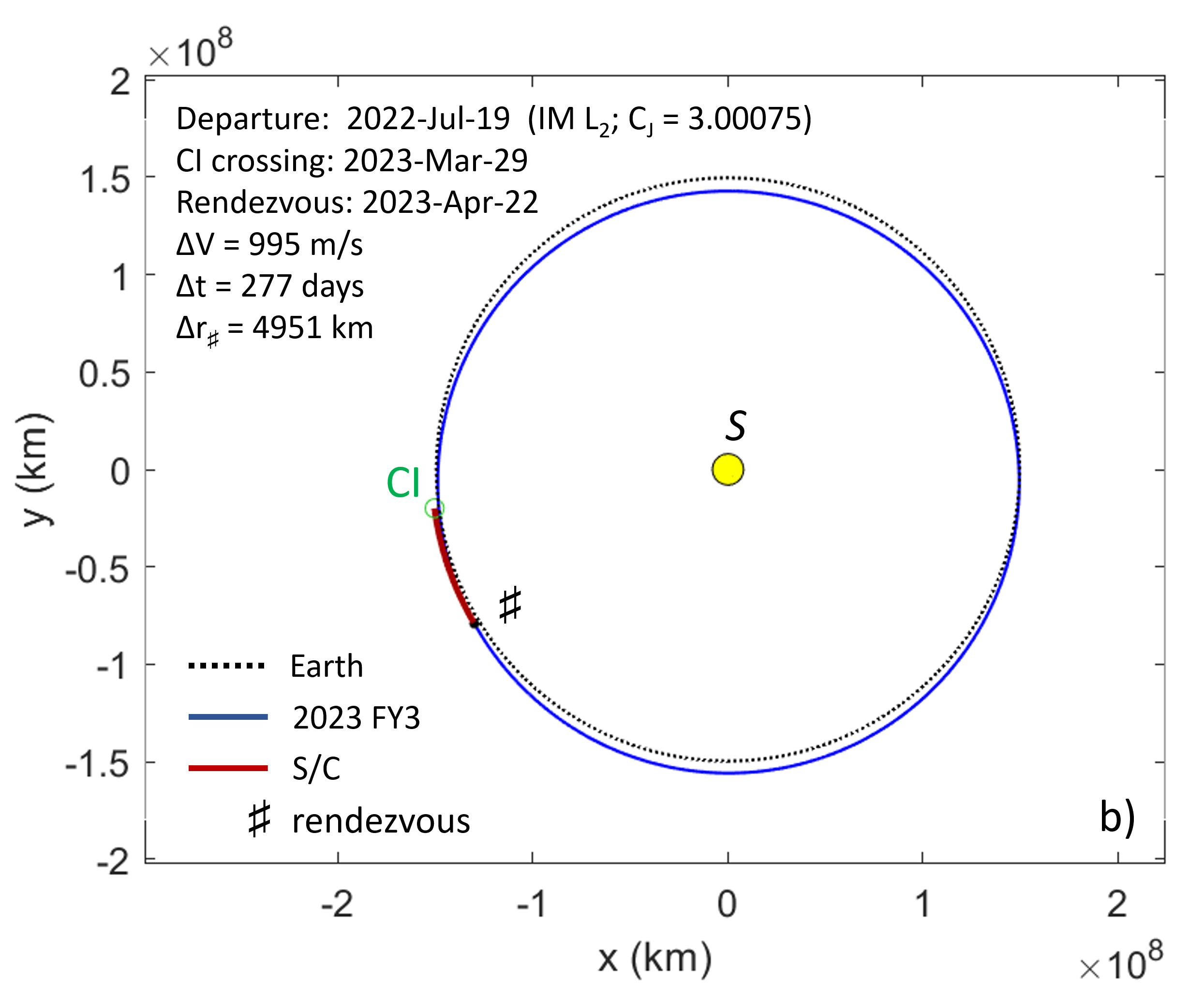} \\
\includegraphics[width=6.1cm]{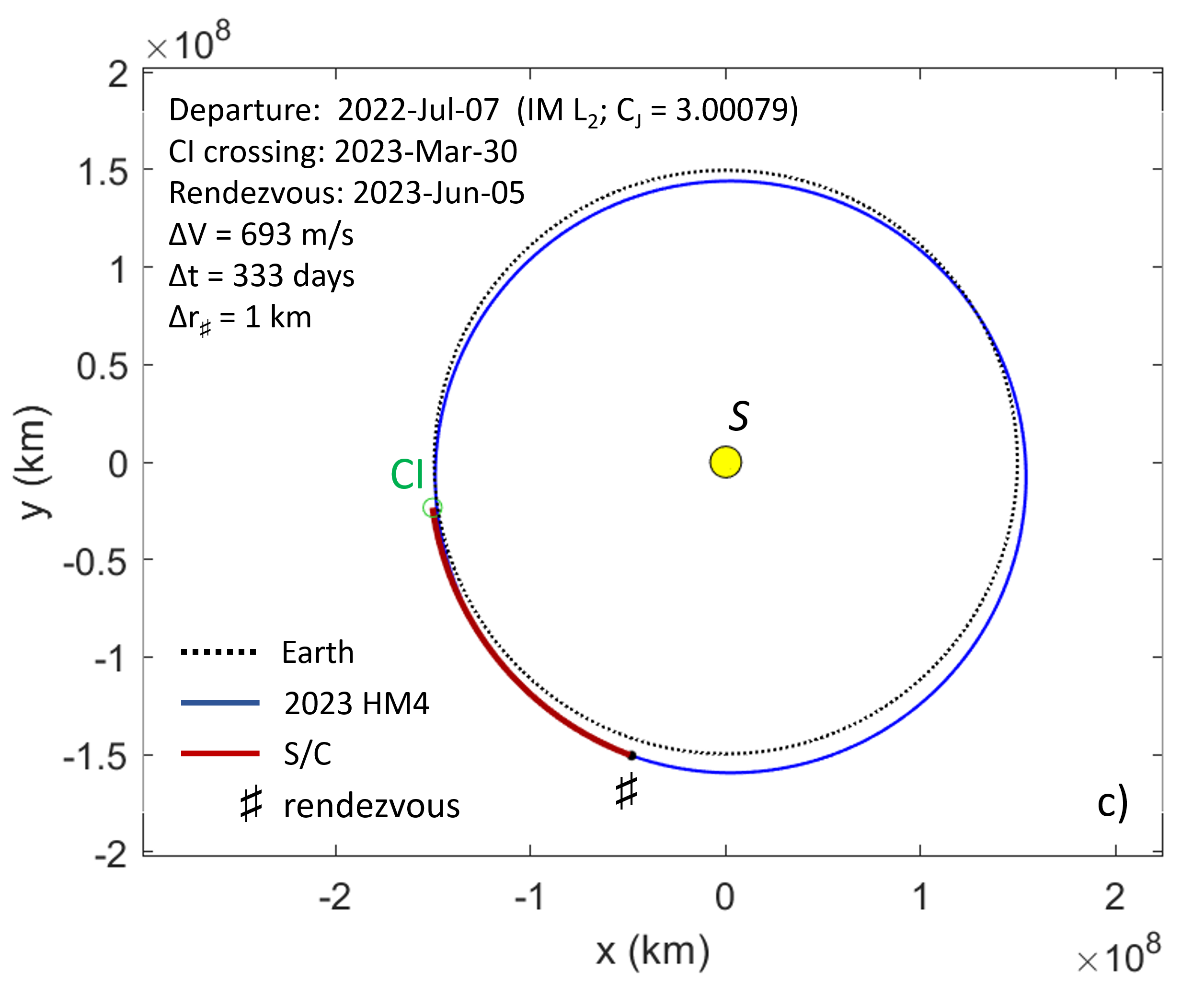} \hspace{-0.3cm}\includegraphics[width=6.1cm]{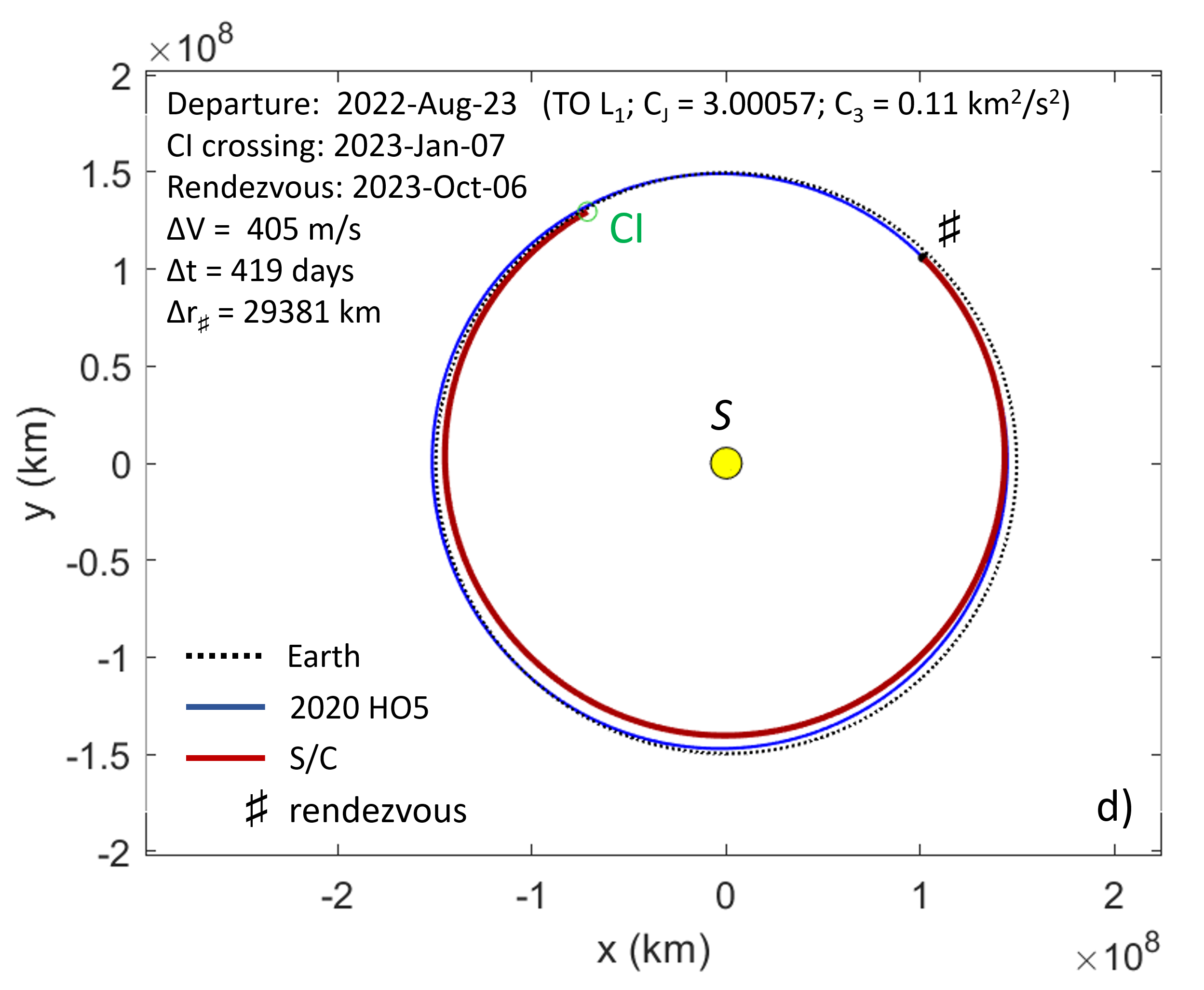} \\
\includegraphics[width=6.1cm]{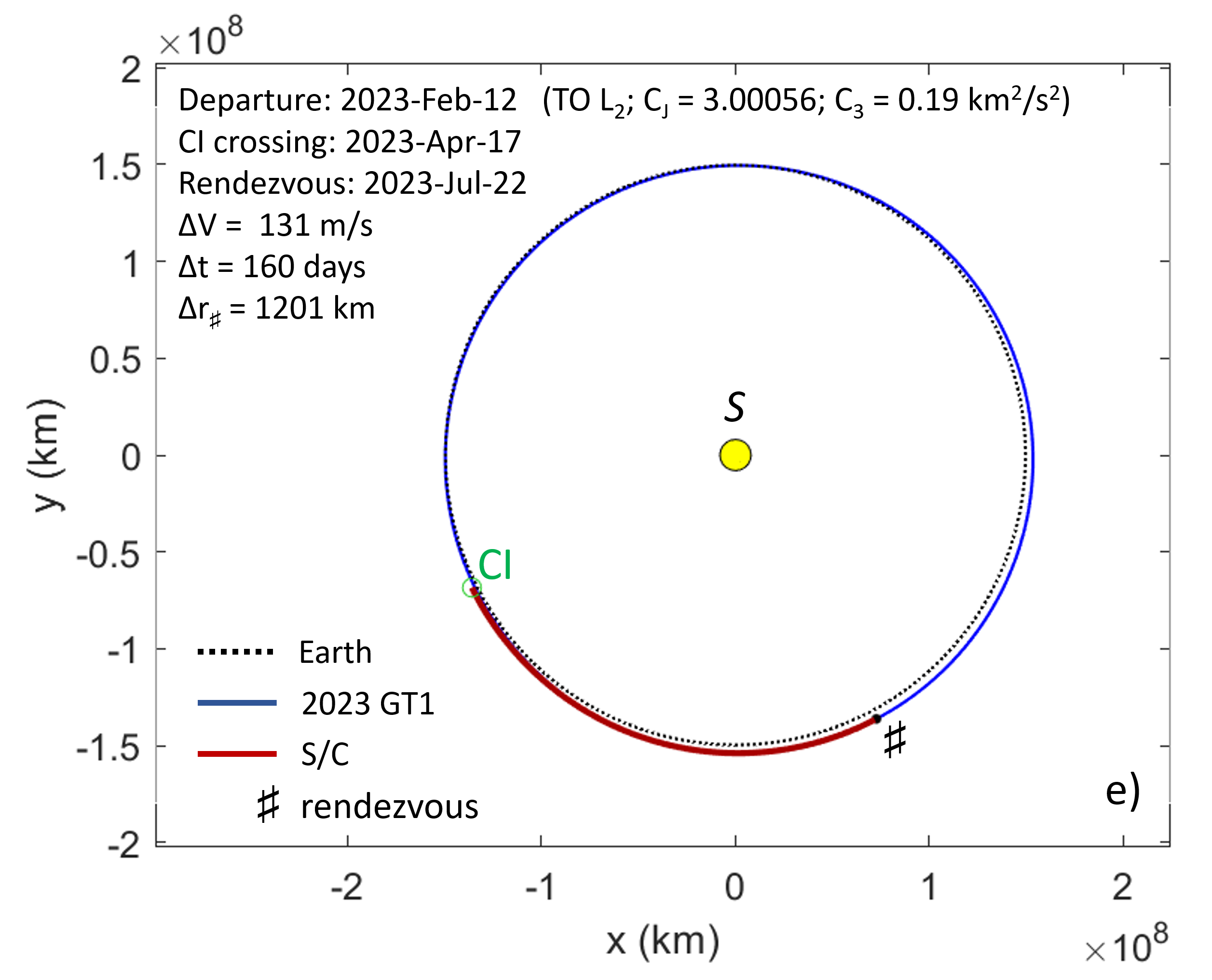}\hspace{-0.3cm} \includegraphics[width=6.1cm]{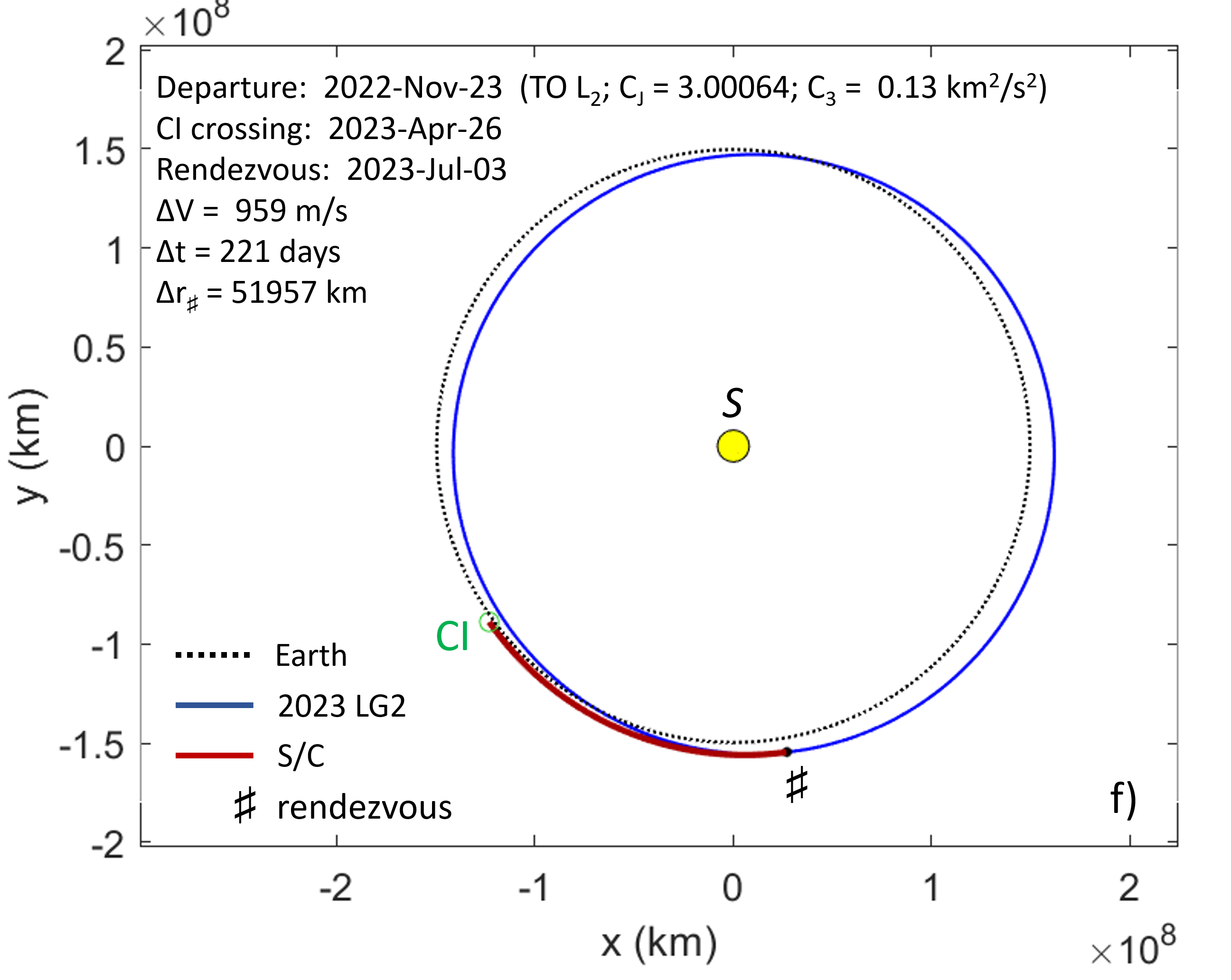} \\
\caption{Examples of planar rendezvous trajectories.}
\label{fig:2Dcases}
\end{figure}

The strategy outlined above has been applied to all the combinations of target orbits and S/C trajectories (MTs and TOs from L$_1$ and L$_2$). The simulations have been conducted with sampling intervals $\Delta \omega$ of 0.4$^{\circ}$ and $\omega_e$ = 0.02$^{\circ}$. Setting an upper limit of 1500 m/s on the magnitude of $\Delta {\bf V_{\#}}$ yields close to 70~000 rendezvous trajectories to a subset of 24 asteroids from the initial list. Table~\ref{tab:2D_sols} lists the number of solutions and the features of the transfers with the best performance for each target and trajectory type: i.e., minimum $\Delta V_{\#}$ ($\Delta V_{\#m}$) and corresponding transfer time, minimum $\Delta t$ ($\Delta t_{m}$) and corresponding rendezvous impulse, and the minimum $\Delta r_{\#}$ ($\Delta r_{\#m}$). Figure~\ref{fig:2Dcases} illustrates six solutions from the set. Table
\ref{tab:lambert} compares the magnitude $\Delta V_{\#}$ of the rendezvous maneuver with the cost of the two-impulse transfer ($\Delta {V}_{CI}$, $\Delta {V}_{\#}'$) for the six cases portrayed in Fig.~\ref{fig:2Dcases}.

\newpage
\begin{landscape}
\begin{table}[h!]
\begin{center}
{\scriptsize
\begin{tabular}{|r|r|rrr|rrr|rrr|rrr|} \hline \hline
& No. of solutions&  & MT-L$_1$  &  & & MT-L$_2$  & & & TO-L$_1$  & & & TO-L$_2$ & \\ 
Ind.  & MT-L$_1$/MT-L$_2$  & $\Delta V_{\#m} (\Delta t)$  & $\Delta t_m /(\Delta V)$ &$\Delta r_{\#m}$ & $\Delta V_{\#m} /(\Delta t)$ & $\Delta t_m /(\Delta V)$ & $\Delta r_{\#m}$ & $\Delta V_{\#m} /(\Delta t)$ & $\Delta t_m /(\Delta V)$ &$\Delta r_{\#m}$ & $\Delta V_{\#m} /(\Delta t)$ & $\Delta t_m /(\Delta V)$ &$\Delta r_{\#m}$  \\
 & /TO-L$_1$/TO-L$_2$ &  m/s (day) &  day (m/s) & km & m/s (day) &  day (m/s)  & km & m/s (day) &  day (m/s) & km & m/s (day) &  day (m/s) & km \\ \hline \hline
   35 &   609/    0/   34/    0
 &   357 (392) & 354 (491) &    82 & - & - & -  &   419 (300) & 250 (543) &  1973 & - & - & - \\ \hline
   39 &   684/    0/    0/    0
 &   716 (554) & 478 (990) &    17 & - & - & -  & - & - & -  & - & - & - \\ \hline
   44 &     0/  882/    0/  198
 & - & - & -  &   344 (402) & 340 (535) &     9 & - & - & -  &   359 (254) & 186 (549) &     5\\ \hline
   45 &     0/ 2981/    0/    0
 & - & - & -  &   378 (666) & 552 (994) &    92 & - & - & -  & - & - & - \\ \hline
   46 &  8042/    0/ 5157/    0
 &   405 (643) & 470 (818) &     8 & - & - & -  &   {\bf 405 (419)} & 310 (997) &     1 & - & - & - \\ \hline
   47 &     0/  330/    0/    0
 & - & - & -  &   676 (675) & 637 (922) &   113 & - & - & -  & - & - & - \\ \hline
   49 &  1488/    0/    7/    0
 &   464 (468) & 399 (921) &    10 & - & - & -  &   572 (334) & 324 (811) &  8905 & - & - & - \\ \hline
   50 &     0/ 2060/    0/    0
 & - & - & -  &   558 (717) & 540 (991) &   134 & - & - & -  & - & - & - \\ \hline
   52 &     0/ 1737/    0/    0
 & - & - & -  &   394 (693) & 604 (995) &     3 & - & - & -  & - & - & - \\ \hline
   54 &  2065/    0/   34/    0
 &   {\bf 332 (612)} & 531 (997) &    12 & - & - & -  &   500 (480) & 463 (702) & 11848 & - & - & - \\ \hline
   55 &  1214/    0/    0/    0
 &   596 (564) & 476 (996) &    16 & - & - & -  & - & - & -  & - & - & - \\ \hline
   56 &     0/ 1025/    0/   10
 & - & - & -  &   585 (550) & 484 (994) &     1 & - & - & -  &   593 (448) & 395 (996) &  2344\\ \hline
   59 &     0/ 1251/    0/   43
 & - & - & -  &   492 (543) & 447 (859) &   161 & - & - & -  &   499 (395) & 356 (750) &   128\\ \hline
   61 &     0/ 3058/    0/  209
 & - & - & -  &   133 (530) & 427 (357) &    34 & - & - & -  &   130 (378) & 301 (271) &    96\\ \hline
   63 &     0/  375/    0/    0
 & - & - & -  &   624 (582) & 498 (996) &   301 & - & - & -  & - & - & - \\ \hline
   64 &     0/ 1076/    0/   24
 & - & - & -  &   133 (513) & 416 (240) &    18 & - & - & -  &   132 (359) & 339 (174) &  1329\\ \hline
   65 &     0/  116/    0/   49
 & - & - & -  &   152 (513) & 510 (164) &   305 & - & - & -  &   168 (354) & 330 (173) &  4724\\ \hline
   66 &     0/ 6233/    0/  814
 & - & - & -  &   432 (397) & {\bf 277 (995)} &     2 & - & - & -  &   447 (152) &  98 (644) &     5\\ \hline
   67 &     0/ 8065/    0/  206
 & - & - & -  &   381 (402) & 299 (814) &    26 & - & - & -  &   599 (268) & 216 (939) &   280\\ \hline
   68 &     0/ 6183/    0/ 5307
 & - & - & -  &   133 (407) & 265 (585) &     7 & - & - & -  &   {\bf 131 (160)} &  90 (566) &    10\\ \hline
   69 &     0/ 4767/    0/   65
 & - & - & -  &   462 (376) & 295 (930) &     {\bf 1} & - & - & -  &   694 (281) & 236 (711) &   129\\ \hline
   70 &     0/ 2562/    0/   84
 & - & - & -  &   244 (381) & 320 (542) &     3 & - & - & -  &   275 (280) & 229 (500) &   731\\ \hline
   71 &     0/ 1560/    0/   49
 & - & - & -  &   548 (411) & 323 (911) &    59 & - & - & -  &   590 (274) & {\bf 221 (959)} &    19\\ \hline
   72 &    15/    0/    0/    0
 &   979 (409) & 398 (997) &   455 & - & - & -  & - & - & -  & - & - & - \\ \hline
\hline
\end{tabular}}
\end{center}
\caption{Characteristics of the best planar rendezvous trajectories to the selected NEAs. The solutions highlighted in boldface are illustrated in Fig.~\ref{fig:2Dcases}.}
\label{tab:2D_sols}
\end{table}
\end{landscape}
\newpage

\begin{table}[t!]
\begin{center}
\begin{tabular}{|c|r|r|r|}
 \hline \hline
Case  & $\Delta V_{\#}$ & $\Delta V_{CI}$ & $\Delta V_{\#}'$ \\
      &   (m/s)          & (m/s)           & (m/s)            \\ \hline \hline 
a     &   332.1            &   0.46             & 332.6  \\ \hline
b     &   995.5            &   20.5             & 992.8  \\ \hline
c     &   693.6            &   1.11             & 694.6    \\ \hline
d     &   405.7            &   0.15             & 405.5  \\ \hline
e     &   131.9           &    0.03            & 131.9   \\ \hline
f     &   959.5            &   10.1             & 962.2  \\ \hline \hline 
\end{tabular}
\end{center}
\caption{Magnitude of the rendezvous maneuver (second column) and cost of the two-impulse transfer (third and fourth column) for the six cases portrayed in Fig.~\ref{fig:2Dcases}.}
\label{tab:lambert}
\end{table}

\section{Extension to 3D impulsive rendezvous trajectories}
\label{sec:3D}
\begin{figure}[h!]
\centering
\includegraphics[width=7.0cm]{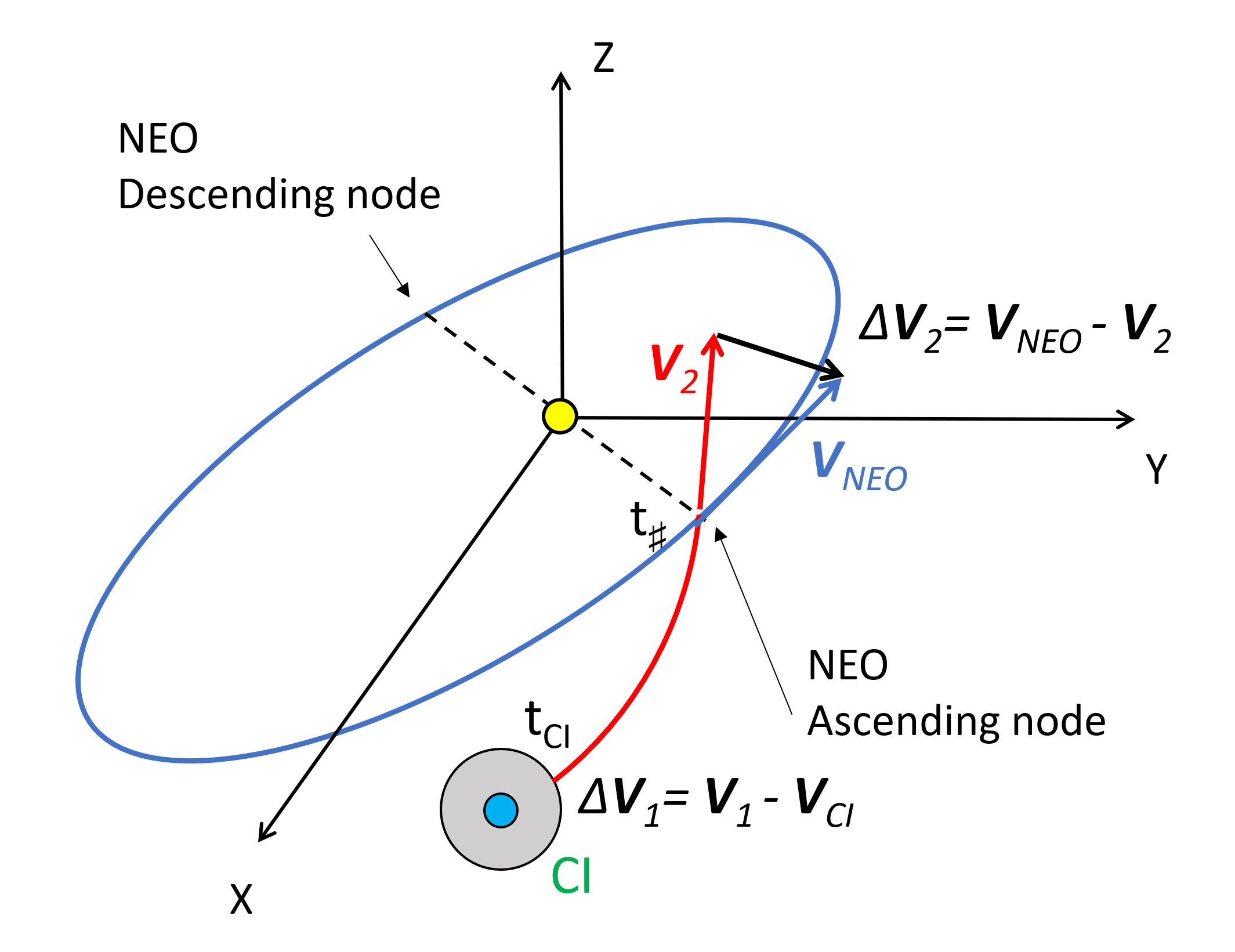}
\caption{Schematic representation of a two-impulse transfer to rendezvous with a target in an inclined orbit. The encounter takes place at one of the nodes.}
\label{fig:3D_Scheme}
\end{figure}

Impulsive rendezvous trajectories taking into account the 3D orientation (inclination, longitude of the ascending node, argument of perihelion) of the orbit of the candidate NEO have been designed by computing the intersections between confocal non-coplanar elliptical orbits, one with fixed orbital elements (NEO) and the other lying on the ecliptic plane and oriented as determined by the date of CI crossing. Under this assumption, the intersection must occur at either node of the target's orbit. However, the correct phasing is not guaranteed \textit{a priori}. The algorithm adjusts the CI crossing date (which determines the departure date) and calculates a ballistic heliocentric arc connecting the CI with a node of the target's orbit. A Lambert solver computes the necessary impulses at the CI ($\Delta {\bf V}_1$) and the rendezvous point ($\Delta {\bf V}_2$) (see Fig.~\ref{fig:3D_Scheme}).

\begin{figure}[h!]
\centering
\includegraphics[width=5.8cm]{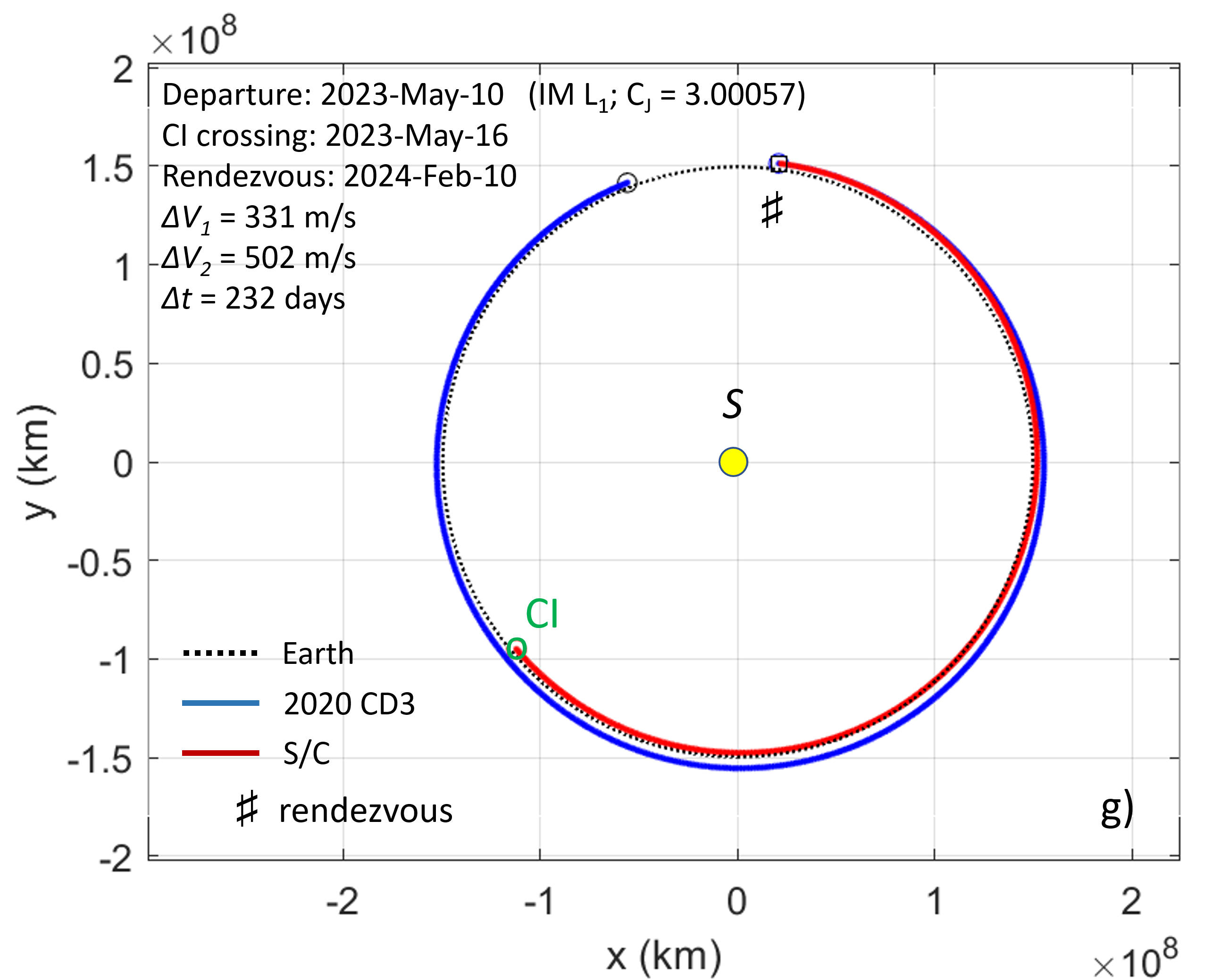} \hspace{-0.3cm} \includegraphics[width=6.2cm]{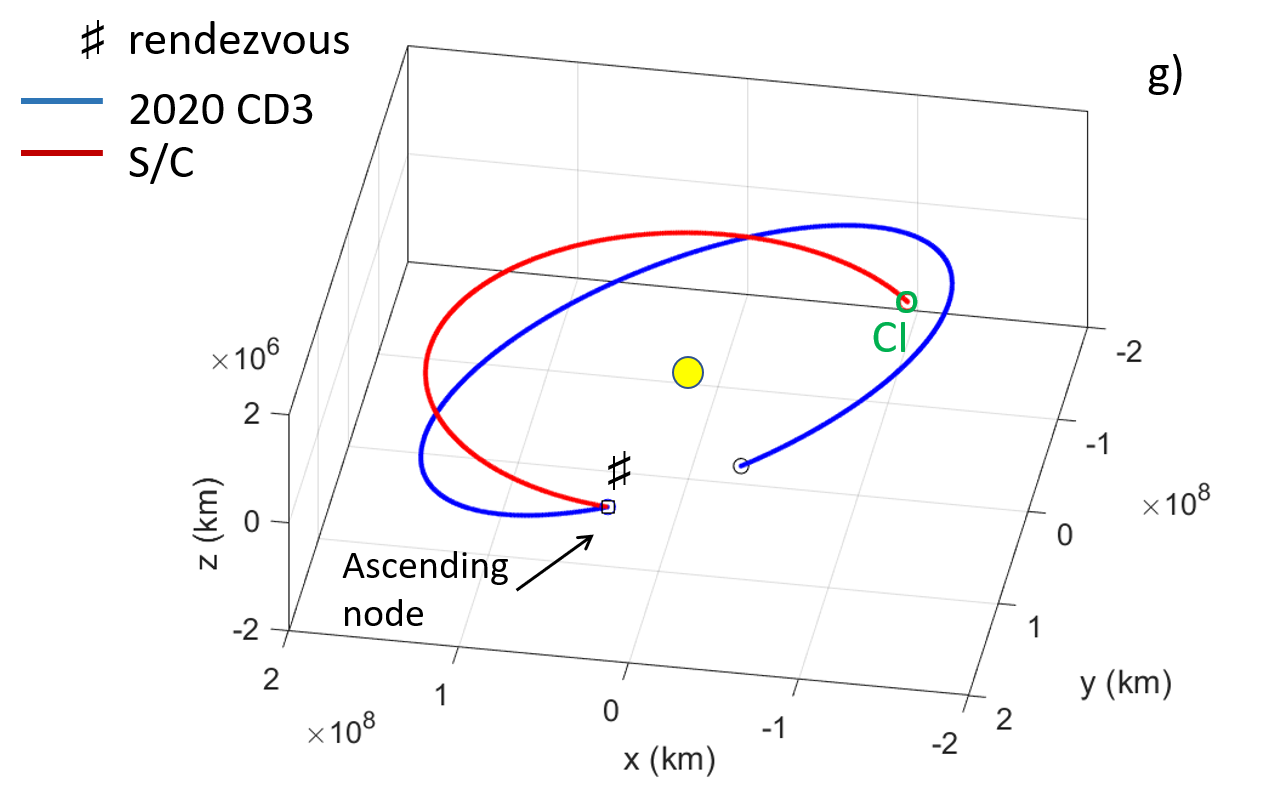} \\
\includegraphics[width=5.8cm]{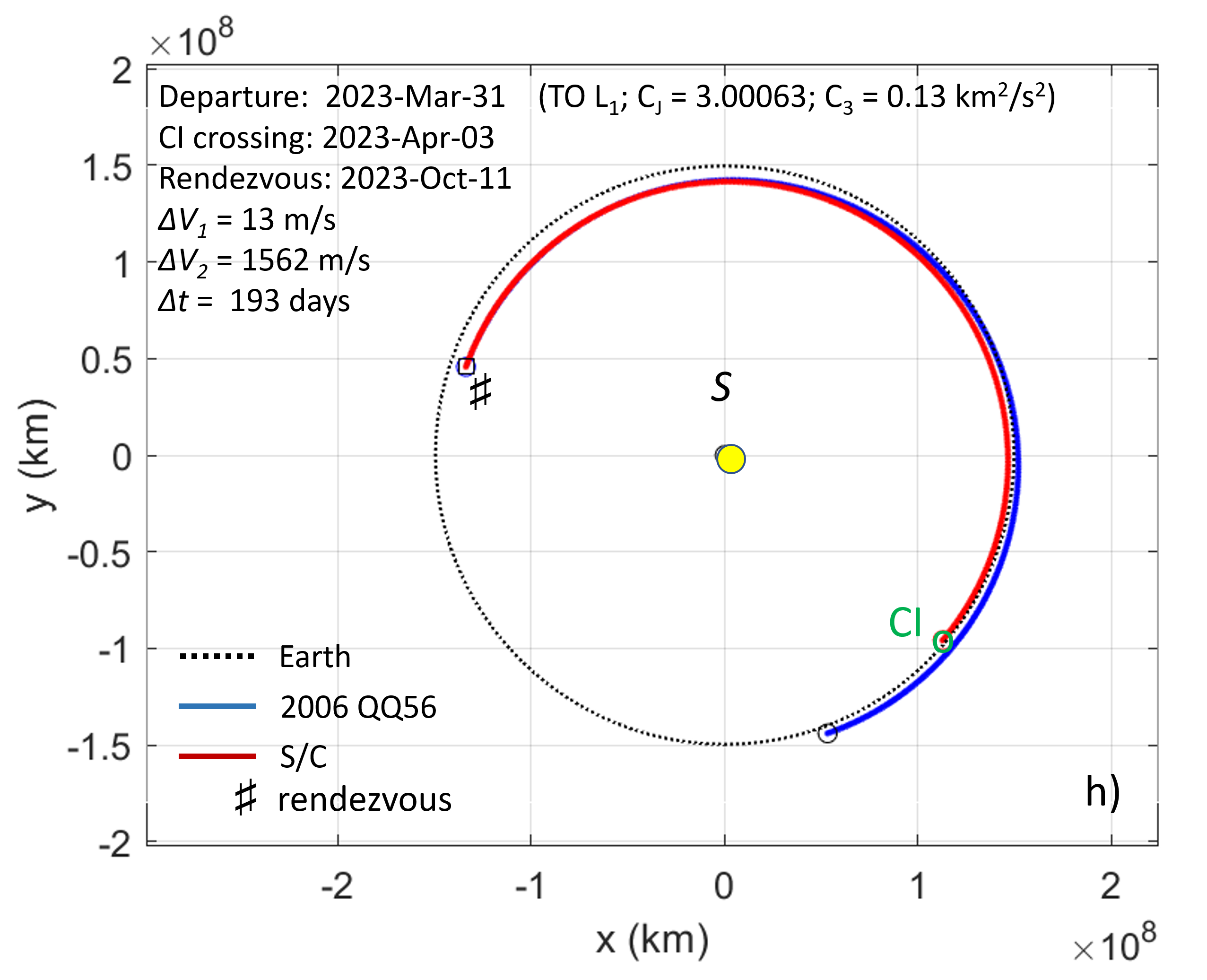} \hspace{-0.3cm} \includegraphics[width=6.2cm]{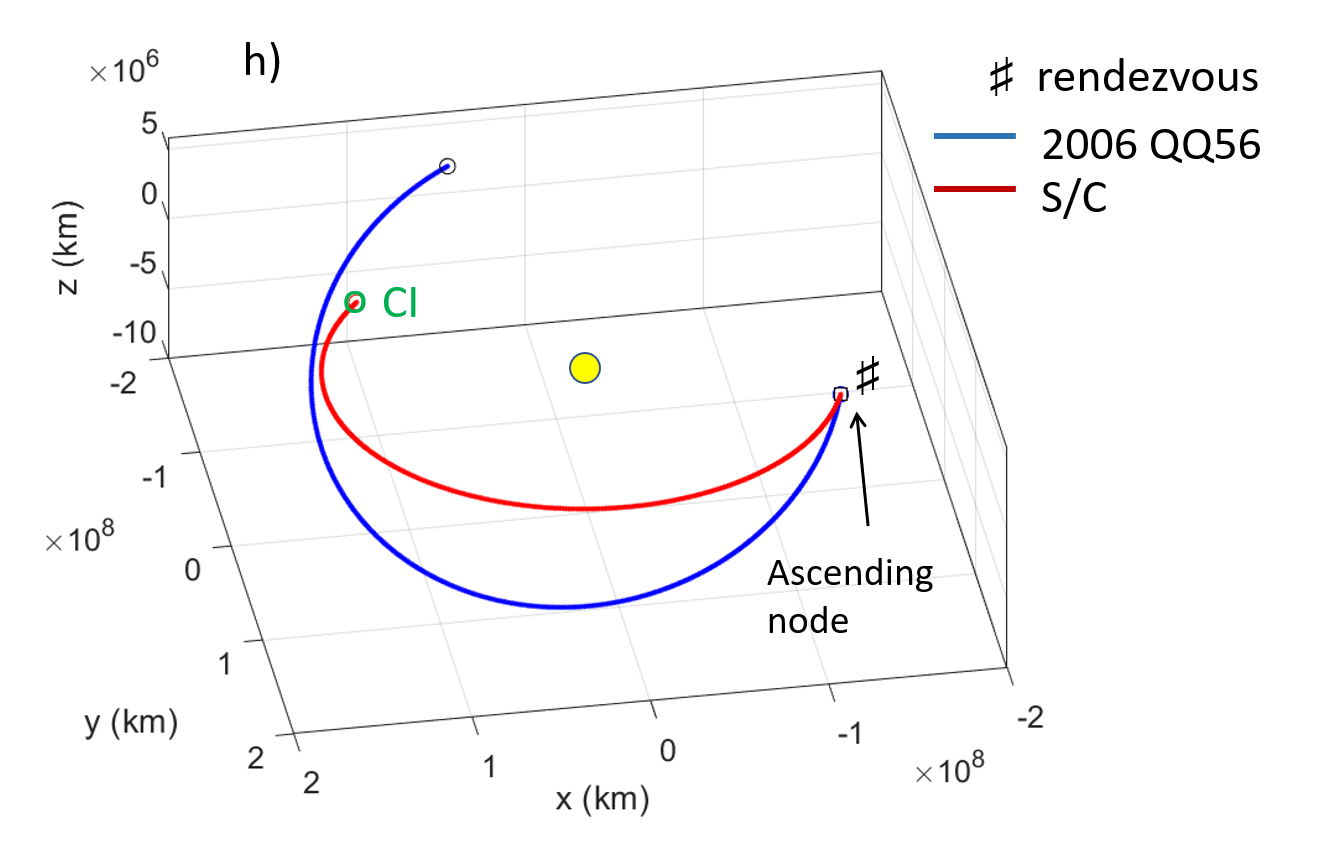} \\
\caption{Example of spatial rendezvous trajectories: ecliptic projection (left) and 3D view (right).}
\label{fig:3Dcases}
\end{figure}

The simulations consider both types of trajectories (MTs and TOs) using a discretization of 5 days for the CI crossing date and setting bounds of 2.5 km/s for the total transfer impulse ($\Delta V = \Delta V_{1} + \Delta V_{2}$) and 1.2 years for the time of flight. The results are summarized in Tables~\ref{tab:3D_sols1} and \ref{tab:3D_sols2}, which for each accessible target and trajectory type yields the number of solutions obtained, minimum $\Delta V$ ($\Delta V_m$) and corresponding transfer time, fastest transfer ($\Delta t_m$) and corresponding $\Delta V$. 
Figure~\ref{fig:3Dcases} illustrates the ecliptic projection and the 3D view of two trajectories.

\newpage

\begin{landscape}
\begin{table}[h!]
\begin{center}
{\scriptsize
\begin{tabular}{|r|r|rr|rr|rr|rr|} \hline \hline
 & No. of solutions &  MT-L$_1$  & & MT-L$_2$  & & TO-L$_1$  & & TO-L$_2$ & \\ 
Index & MT-L$_1$/MT-L$_2$ & $\Delta V_m (\Delta t)$  & $\Delta t_m /(\Delta V)$  & $\Delta V_m /(\Delta t)$ & $\Delta t_m /(\Delta V)$ & $\Delta V_m /(\Delta t)$ & $\Delta t_m /(\Delta V)$  & $\Delta V_m /(\Delta t)$ & $\Delta t_m /(\Delta V)$   \\
 & /TO-L$_1$/TO-L$_2$ &  m/s (day) &  day (m/s) & m/s (day) &  day (m/s) & m/s (day) &  day (m/s) & m/s (day) &  day (m/s)  \\ \hline \hline
    2 &  2878/ 5254/    0/    0
 & 1189 ( 88) &   11 (1558) &  769 (136) &   65 (1924) & - & -  & - & - \\ \hline
    3 &  2349/ 2829/    0/    0
 &  941 (249) &   97 (2463) & 1284 (273) &   46 (2443) & - & -  & - & - \\ \hline
    4 &     0/    0/   84/    0
 & - & -  & - & -  & 2321 (346) &   49 (2443) & - & - \\ \hline
    5 &  1349/ 2172/    0/    0
 & 2161 (344) &  158 (2495) & 2188 (337) &   57 (2498) & - & -  & - & - \\ \hline
    6 &     0/    0/   61/   54
 & - & -  & - & -  & 1575 (193) &  116 (2069) & 1679 (296) &   43 (2208)\\ \hline
    7 &     0/    0/  983/ 1077
 & - & -  & - & -  &  853 (204) &  182 ( 971) &  574 (184) &   45 ( 782)\\ \hline
    8 &   692/ 1810/    0/    0
 & 2263 (326) &  180 (2482) & 2078 (313) &  181 (2463) & - & -  & - & - \\ \hline
    9 &     0/    0/  275/  616
 & - & -  & - & -  & 1331 (275) &   11 (2009) &  965 (294) &  207 (2175)\\ \hline
   10 &     0/    1/   46/   47
 & - & -  & 2320 ( 98) &   98 (2320) & 1483 (114) &   83 (2012) & 2153 (136) &  136 (2153)\\ \hline
   11 &  1159/    3/    0/    0
 & 1904 (315) &   10 (2365) & 1692 ( 39) &   33 (2347) & - & -  & - & - \\ \hline
   12 &     0/    0/   32/  127
 & - & -  & - & -  & 2237 (209) &  124 (2375) & 1545 (219) &  131 (2378)\\ \hline
   13 &     0/    0/    0/  189
 & - & -  & - & -  & - & -  & 1817 ( 55) &   55 (1817)\\ \hline
   14 &     0/    0/  100/  144
 & - & -  & - & -  &  640 (122) &   89 ( 753) & 1168 (333) &  186 (1562)\\ \hline
   16 &     0/    0/    8/   79
 & - & -  & - & -  & 2351 (221) &  175 (2470) & 1658 (289) &   58 (2388)\\ \hline
   17 &     0/    0/  427/  506
 & - & -  & - & -  & 1313 (134) &  130 (1330) &  737 (288) &   93 (2190)\\ \hline
   18 &     0/    0/  752/  840
 & - & -  & - & -  & 1503 (260) &  239 (1673) & 1323 (245) &   96 (2493)\\ \hline
   19 &     0/    0/  151/  243
 & - & -  & - & -  & 2275 (245) &  220 (2348) & 1949 (360) &  125 (2357)\\ \hline
   20 &  1952/ 2880/    0/    0
 & 1433 (280) &  162 (1591) & 1157 ( 71) &   22 (1446) & - & -  & - & - \\ \hline
   21 &     0/    0/   46/   21
 & - & -  & - & -  & 2028 (150) &  119 (2438) & 2014 (151) &  150 (2285)\\ \hline
   22 &     0/    0/  135/  415
 & - & -  & - & -  & 2429 (172) &  166 (2434) & 2150 ( 27) &   19 (2321)\\ \hline
   23 &  2835/ 1947/    0/    0
 & 1496 ( 31) &   15 (2180) &  897 ( 97) &   62 (2392) & - & -  & - & - \\ \hline
   24 &  2137/    0/    0/    0
 & 1567 (331) &   82 (2289) & - & -  & - & -  & - & - \\ \hline
   25 &     0/    0/   41/   89
 & - & -  & - & -  & 2200 (181) &  167 (2283) & 1460 (208) &  100 (2305)\\ \hline
   26 &   519/  971/    0/    0
 & 1798 (122) &  102 (1821) & 1720 (253) &   67 (2353) & - & -  & - & - \\ \hline
   28 &     0/  724/    5/    0
 & - & -  & 1215 ( 56) &   44 (2120) & 2327 (280) &  275 (2474) & - & - \\ \hline
   29 &     0/   78/    0/    0
 & - & -  & 1978 ( 41) &   34 (2434) & - & -  & - & - \\ \hline
   30 &     0/    0/    2/    0
 & - & -  & - & -  & 2424 (310) &  305 (2439) & - & - \\ \hline
\hline
\end{tabular}}
\end{center}
\caption{Characteristics of the best 3D rendezvous trajectories to the selected NEAs with $\Delta V \le 2500$ m/s and $\Delta t  \le 1.2$ years.}
\label{tab:3D_sols1}
\end{table}

\newpage
\begin{table}[h!]
\begin{center}
{\scriptsize
\begin{tabular}{|r|r|rr|rr|rr|rr|} \hline \hline
 & No. of solutions &  MT-L$_1$  & & MT-L$_2$  & & TO-L$_1$  & & TO-L$_2$ & \\ 
Index & MT-L$_1$/MT-L$_2$ & $\Delta V_m (\Delta t)$  & $\Delta t_m /(\Delta V)$  & $\Delta V_m /(\Delta t)$ & $\Delta t_m /(\Delta V)$ & $\Delta V_m /(\Delta t)$ & $\Delta t_m /(\Delta V)$  & $\Delta V_m /(\Delta t)$ & $\Delta t_m /(\Delta V)$   \\
 & /TO-L$_1$/TO-L$_2$ &  m/s (day) &  day (m/s) & m/s (day) &  day (m/s) & m/s (day) &  day (m/s) & m/s (day) &  day (m/s)  \\ \hline \hline
   31 &  3788/ 3578/    0/    0
 & 1610 (338) &  158 (1881) &  929 ( 91) &   59 (2342) & - & -  & - & - \\ \hline
   32 &     0/    0/  220/  164
 & - & -  & - & -  & 1833 (180) &  148 (2116) & 1633 (286) &  176 (2489)\\ \hline
   34 &     0/    0/    0/    2
 & - & -  & - & -  & - & -  & 2471 (217) &  217 (2474)\\ \hline
   36 &   779/  473/    0/    0
 & 1805 (192) &  190 (2041) & 1211 (158) &   94 (2488) & - & -  & - & - \\ \hline
   37 &  2911/ 5761/    0/    0
 & 1575 (210) &  203 (2250) &  856 (225) &  139 (1754) & - & -  & - & - \\ \hline
   38 &     0/ 3325/    0/    0
 & - & -  & 1898 (123) &   63 (2490) & - & -  & - & - \\ \hline
   39 &     0/    0/   27/  157
 & - & -  & - & -  & 2392 (332) &  332 (2392) & 2188 (358) &  358 (2216)\\ \hline
   40 &   533/ 3670/    0/    0
 & 2329 (166) &  148 (2467) & 1608 (109) &   71 (2366) & - & -  & - & - \\ \hline
   41 &  1574/ 2204/    0/    0
 & 1427 (123) &  123 (1427) &  802 ( 59) &   43 (2051) & - & -  & - & - \\ \hline
   42 &   591/  889/    0/    0
 &  833 (232) &  165 ( 876) &  796 (202) &   58 (1636) & - & -  & - & - \\ \hline
   43 &     3/  353/    0/    0
 & 2315 (182) &  182 (2315) & 1924 (260) &  152 (2428) & - & -  & - & - \\ \hline
   44 &     0/    0/    1/    3
 & - & -  & - & -  & 2419 ( 29) &   29 (2419) & 2126 ( 41) &   29 (2133)\\ \hline
   48 &    45/  769/    0/    0
 & 2146 (170) &  162 (2387) & 1566 (267) &   23 (2196) & - & -  & - & - \\ \hline
   50 &   200/  234/    0/    0
 & 1180 (129) &  129 (1180) & 1419 (293) &  267 (1925) & - & -  & - & - \\ \hline
   53 &     1/  107/    0/    0
 & 2485 (225) &  225 (2485) & 2100 (160) &   58 (2459) & - & -  & - & - \\ \hline
   54 &     0/    0/    1/  416
 & - & -  & - & -  & 2472 (343) &  343 (2472) & 1568 (326) &  323 (1616)\\ \hline
   55 &     0/    0/    0/    1
 & - & -  & - & -  & - & -  & 2324 ( 23) &   23 (2324)\\ \hline
   56 &     2/   22/    5/    0
 & 1892 (338) &  336 (2337) & 1945 (164) &  140 (2477) & 1566 ( 70) &   36 (2018) & - & - \\ \hline
   59 &     0/    3/    0/    0
 & - & -  & 2449 (218) &  218 (2456) & - & -  & - & - \\ \hline
   61 &    11/    0/    0/    0
 & 1762 (303) &  290 (2439) & - & -  & - & -  & - & - \\ \hline
   62 &     0/    0/    1/    1
 & - & -  & - & -  & 2435 ( 34) &   34 (2435) & 2219 ( 32) &   32 (2219)\\ \hline
   64 &     0/    0/    0/    1
 & - & -  & - & -  & - & -  & 1983 ( 23) &   23 (1983)\\ \hline
   65 &     0/    0/    0/    2
 & - & -  & - & -  & - & -  & 2078 ( 27) &   27 (2078)\\ \hline
   66 &     1/   63/    1/    2
 & 2481 (307) &  307 (2481) & 1840 (348) &  334 (2495) & 2143 ( 33) &   33 (2143) & 1356 ( 53) &   29 (1543)\\ \hline
   67 &     0/    0/    0/    1
 & - & -  & - & -  & - & -  & 2297 ( 24) &   24 (2297)\\ \hline
   69 &     1/    0/    0/    0
 & 2352 (280) &  280 (2352) & - & -  & - & -  & - & - \\ \hline
   70 &     0/    0/    0/    1
 & - & -  & - & -  & - & -  & 2050 ( 21) &   21 (2050)\\ \hline
   71 &     0/    0/    0/    2
 & - & -  & - & -  & - & -  & 2335 ( 60) &   60 (2335)\\ \hline
\hline
\end{tabular}}
\end{center}
\caption{Characteristics of the best 3D rendezvous trajectories to the selected NEAs  with $\Delta V \le 2500$ m/s and $\Delta t  \le 1.2$ years.}
\label{tab:3D_sols2}
\end{table}
\end{landscape}
\newpage

\section{Comparison with patched-conics solutions}
\label{sec:comp}
The performance of the 3D low-energy impulsive transfers via TOs described in this contribution have been compared with the trajectories of the mission concept study carried out within NASA's Planetary Science Decadal Survey devoted to NEAs \cite{Strange:2010}. It provides direct, HT rendezvous trajectories with launch dates between 1-Jan-2020 and 1-Jan-2025, maximum flight time to the target of 5 years, launch C$_3$ below 25 km$^2$/s$^2$, and rendezvous $\Delta V$ under 3 km/s. NASA's solutions are computed using a simplified, suboptimal, patched-conics approximation and then optimized with respect to the delivered S/C mass at rendezvous via Sequential Quadratic Programming. The work of Strange et al. \cite{Strange:2010} was published in 2010, when the majority of the asteroids considered in this work were yet to be discovered. As a result, the intersection of the two datasets contains only two targets: 2003 YN107 (object index 4) and 2006 QQ56 (object index 6). Table~\ref{tab:two_sols} compares total transfer $\Delta V$, time of flight $\Delta t$ and launch C$_3$ from Appendix C of \cite{Strange:2010} against the 3D TO/L$_1$ transfers with the lowest $\Delta V$ (see Table~\ref{tab:3D_sols1}) to the same targets. Note that the solutions by Strange et al. \cite{Strange:2010} launch approximately 2.5 years earlier than those obtained in this work.

The cost of the rendezvous with 2003 YN107 is significantly higher when traveling via a TO (2321 m/s versus 1250 m/s), but this disadvantage is partially offset by a lower launch energy. Assuming a 300-km altitude circular departure orbit, the two values of the launch energy (0.13 km$^2$/s$^2$ and 6.30 km$^2$/s$^2$) map into departure $\Delta V$'s of 3212 m/s and 3490 m/s. Thus, the net advantage of NASA's solution is 793 m/s. Note, however, that this comes at the expense of a much longer transfer duration (almost double). It is generally possible to reduce propellant consumption by increasing the time of flight, so it is not surprising that the longer transfer is more efficient. Note also that the search for feasible trajectories limited the time of flight to 1.2 years. The limit made it impossible to find a close match of NASA's solution, but it is not an intrinsic limitation of the methodology. Raising the maximum time of flight for the search enables revisiting the intersection points of the heliocentric ellipses once per orbital period of the S/C. Thus, additional phasing conditions can be analyzed for improved fuel-efficiency. This enables a comprehensive exploration of the solution space, making the technique quite versatile.

The trajectory to 2006 QQ56 proposed in this work (see also case h, Fig.~\ref{fig:3Dcases}) outperforms the solution by \cite{Strange:2010} in all aspects (shorter duration and lower total impulse). This demonstrates the ability of the new technique to find highly-efficient solutions, even when the time of flight is constrained. It is a valuable tool to expand the solution space covered by more traditional methods.

\begin{table}[h!]
\begin{center}
\begin{tabular}{|l|r|r|r|r|}
\hline \hline 
Object &  \multicolumn{2}{c|}{2003 YN107 (index 4)} & 
    \multicolumn{2}{c|}{2006 QQ56 (index 6)}\\ \hline
Source		& Ref. \cite{Strange:2010} & This work  & Ref. \cite{Strange:2010} & This work \\ \hline 
$\Delta V$ (m/s) &  1250 & 2321 & 1885 & 1575 \\ \hline
$\Delta t$ (day) &  621 & 346 & 621 & 193 \\ \hline
C$_3$ (km$^2$/s$^2$) & 6.30 & 0.13 & 5.90 & 0.13\\ \hline
Launch date & 2019-Dec-19 & 2022-Mar-26 & 2020-Sep-02 & 2023-Mar-31 \\ \hline
Arrival date & 2021-Aug-26 & 2023-Mar-06 & 2022-May-07 & 2023-Oct-11 \\ \hline \hline
\end{tabular}
\end{center}
\caption{Performance features of the direct rendevouz trajectories to the same two targets computed by Strange et al. \cite{Strange:2010} and in this work.}
\label{tab:two_sols}
\end{table}

\section{Discussion and conclusions}
\label{sec:concl} 
The present contribution proposes a simple and efficient technique to systematically generate low-energy impulsive rendezvous trajectories to Near-Earth Objects on low-eccentricity, low-inclination orbits. The dynamical model is the patched Sun-Earth circular restricted three-body problem / Sun-spacecraft two-body problem. Hence, the influence of the Sun is incorporated along the entire path. Two types of initial conditions are considered: departures from planar Lyapunov orbits around Sun-Earth L$_1$ or L$_2$ via outward branches of the associated hyperbolic invariant manifolds, and launches from a 300-km altitude circular orbit via selected transit orbits (trajectories internal to the same hyperbolic invariant manifolds). Thus, two mission scenarios are accommodated: direct Earth departures and transfers from an intermediate gateway near a Lagrange point. The two types of trajectories exhibit different performance characteristics, particularly in terms of time of flight. Earth departures are generally faster due to their specific geometrical characteristics.

The heliocentric osculating two-body orbit obtained by transforming the state of the spacecraft at the crossing of Earth's sphere of influence lies on the ecliptic plane. The mean motion of the orbit of the Earth around the Sun along with the departure date determine the longitude of the perihelion of such orbit, providing a fundamental degree of freedom (the optimization variable) of the method, which calculates geometrical intersections between the elliptical orbits of the spacecraft and the target (in the planar case, the latter is projected on the ecliptic). Thus, when the rendezvous conditions are met, the transfer requires a single impulsive maneuver at arrival, the cost of which depends on the choice of the intersection point used for the encounter. The extension of the method to inclined target orbits introduces a Lambert arc between the sphere of influence of the Earth and one of the nodes of the orbit of the target, requiring two impulses.   

In 2D, the identification of the rendezvous conditions is an analytical problem that can be solved in closed form. The $\Delta V$-optimal transfer automatically emerges from the variation of the departure date as part of the search for the rendezvous conditions. The simple dependence of the orbital elements of the heliocentric orbit of the spacecraft on the departure state and date helps reduce the computational burden. Furthermore, the propagation of transit orbits and invariant manifold trajectories to the boundary of the sphere of influence of the Earth needs to be carried out only once. All these features make the method computationally efficient. The entire simulation for 72 candidate targets using the full set of invariant manifolds and transit orbits through both libration points took approximately 40 minutes in a mid-range laptop when programmed in the Fortran language.

It is important to note that the method here described identifies all the direct rendezvous opportunities that exist in the selected conditions (resolution, tolerances and maximum allowed $\Delta V$ budget and transfer time) at the chosen energy levels and in the approximation offered by the dynamical model. Hence, an extension to include other types of departure conditions would lead to a general tool capable of generating optimal direct interplanetary trajectories. 

The patched three-body/two-body approximation adopted is at least as accurate as the conventional patched-conics approach. Hence, it is suitable for feasibility analyses, and its results can be refined with an ephemeris model for higher-fidelity mission design. The impulsive maneuvers can be replaced with low-thrust arcs without altering the transfer times (see \cite{CanalesIAC:2023}), reducing the propellant consumption due to the higher specific impulse. The direct rendezvous trajectories described in this contribution can be viewed also as the building blocks of more complex mission scenarios, such as sample return operations and asteroid tours. Furthermore, the $\Delta V$ budgets and transfer times easily meet with the requirements set by NASA on the search for targets accessible by crewed missions \cite{Abell:2012}.

The 3D trajectories from Earth to two neighboring asteroids via transit orbits have been benchmarked against direct optimal patched-conics transfers from a NASA study of accessible targets \cite{Strange:2010}. The test demonstrated the potential of the new method to find solutions that compare favorably in terms of time or flight and/or propellant consumption relative to traditional approaches, while requiring only modest computational resources.

\section*{Acknowledgments}
EF, RF and GD acknowledge Khalifa University of Science and Technology's internal grant CIRA-2021-65/8474000413. EF and RF have been partially supported by Technology Innovation Institute through grant ELLIPSE/8434000533. EF acknowledges projects PID2020-112576GB-C21 and PID2021-123968NB-I00 of the Spanish Ministry of Science and Innovation.


\bibliography{References}

\end{document}